\documentclass[10pt,letterpaper]{article}

\usepackage[margin=1in]{geometry}
\usepackage[T1]{fontenc}
\usepackage{times}
\usepackage{booktabs}
\usepackage{xcolor}

\usepackage{enumitem}
\usepackage{algorithm}
\usepackage{algpseudocode}
\usepackage{colortbl}
\usepackage{tabularx}
\usepackage{graphicx}
\usepackage{microtype}
\usepackage{multirow}
\usepackage[authoryear,round]{natbib}
\setcitestyle{citesep={;},aysep={,},yysep={;}}
\makeatletter
\renewcommand\normalsize{%
  \@setfontsize\normalsize{10}{12}%
  \abovedisplayskip 6pt plus 2pt minus 2pt
  \belowdisplayskip 6pt plus 2pt minus 2pt
  \abovedisplayshortskip 3pt plus 1pt
  \belowdisplayshortskip 4pt plus 1pt minus 1pt
  \let\@listi\@listI
}
\normalsize
\renewcommand\section{\@startsection{section}{1}{\z@}%
  {-11pt plus -2pt minus -2pt}{6pt plus 1pt}%
  {\normalfont\large\scshape\raggedright}}
\renewcommand\subsection{\@startsection{subsection}{2}{\z@}%
  {-9pt plus -2pt minus -1pt}{4pt plus 1pt}%
  {\normalfont\normalsize\scshape\raggedright}}
\renewcommand\subsubsection{\@startsection{subsubsection}{3}{\z@}%
  {-8pt plus -1pt minus -1pt}{3pt plus 1pt}%
  {\normalfont\normalsize\bfseries\raggedright}}
\renewcommand\paragraph{\@startsection{paragraph}{4}{\z@}%
  {6pt plus 1pt minus 1pt}{-1em}{\normalfont\normalsize\bfseries}}
\renewcommand\@maketitle{%
  \begin{center}
    {\fontsize{16}{18}\selectfont\bfseries\@title\par}%
    \vskip 10pt
    {\normalsize\@author\par}%
  \end{center}
  \vskip 3pt
}
\long\def\@makecaption#1#2{%
  \vskip\abovecaptionskip
  \begingroup\small
  \sbox\@tempboxa{#1: #2}%
  \ifdim\wd\@tempboxa>\hsize
    #1: #2\par
  \else
    \global\@minipagefalse
    \hb@xt@\hsize{\hfil\box\@tempboxa\hfil}%
  \fi
  \endgroup
  \vskip\belowcaptionskip
}
\makeatother
\AtBeginDocument{%
  \setlength{\abovedisplayskip}{6pt plus 2pt minus 2pt}%
  \setlength{\belowdisplayskip}{6pt plus 2pt minus 2pt}%
  \setlength{\abovedisplayshortskip}{3pt plus 1pt}%
  \setlength{\belowdisplayshortskip}{4pt plus 1pt minus 1pt}%
  \setlength{\bibsep}{2pt plus 1pt}%
}

\usepackage{amsmath,amsfonts,bm}

\def\eqref#1{equation~\ref{#1}}

\def\1{\bm{1}}

\DeclareMathAlphabet{\mathsfit}{\encodingdefault}{\sfdefault}{m}{sl}
\SetMathAlphabet{\mathsfit}{bold}{\encodingdefault}{\sfdefault}{bx}{n}

\usepackage{xurl}
\usepackage[hidelinks]{hyperref}
\hypersetup{
  bookmarksnumbered=true,
  pdftitle={Spatial Audio Generation and Editing with Executable Plans and Self-Distilling Transfusion},
  pdfauthor={Kam Man Wu; Yu Zhang; Pengjun Fang; Mingfei Chen; Weijia Chen; Zeyue Tian; Qifeng Chen},
  pdfsubject={Spatial audio generation and editing},
  pdfkeywords={AMBIT, spatial audio, Ambisonics, audio generation, audio editing, self-distillation}
}

\AtBeginDocument{%
  \addtocontents{toc}{\protect\setcounter{tocdepth}{-1}}%
}

\usepackage{fancyhdr}

\fancypagestyle{plain}{%
  \fancyhf{}
  \fancyhead[L]{\normalfont\normalsize Preprint}
  \fancyfoot[C]{\thepage}

}

\renewcommand{\topfraction}{0.95}
\renewcommand{\bottomfraction}{0.8}
\renewcommand{\textfraction}{0.05}
\renewcommand{\floatpagefraction}{0.8}

\definecolor{sceneplanblue}{RGB}{33,102,172}
\definecolor{semanticgreen}{RGB}{35,139,69}
\definecolor{spatialorange}{RGB}{230,85,13}
\definecolor{latentpurple}{RGB}{117,107,177}

\newcommand{\method}{\textsc{Ambit}}
\newcommand{\sceneplan}{\textsc{ScenePlan}}

\newcolumntype{Y}{>{\raggedright\arraybackslash}X}
\newcolumntype{C}[1]{>{\centering\arraybackslash}p{#1}}

\title{Spatial Audio Generation and Editing with Executable Plans and Self-Distilling Transfusion}
\author{%
  \begin{tabular}{c}
    Kam Man Wu\textsuperscript{*1,2}\quad
    Yu Zhang\textsuperscript{*}\quad
    Pengjun Fang\textsuperscript{1,2}\quad
    Mingfei Chen\textsuperscript{3}\\[4pt]
    Weijia Chen\textsuperscript{2}\quad
    Zeyue Tian\textsuperscript{$\dagger$1,2}\quad
    Qifeng Chen\textsuperscript{$\dagger$1}\\[8pt]
    \small\textsuperscript{1}HKUST\qquad
    \small\textsuperscript{2}Noiz AI\qquad
    \textsuperscript{3}University of Washington\\[6pt]
    \footnotesize\textsuperscript{*}Equal contribution.\qquad
    \textsuperscript{$\dagger$}Corresponding authors.
  \end{tabular}%
}

\date{}

\begin{document}
\maketitle

\begin{abstract}
Spatial audio authoring coordinates source decisions with soundfield
synthesis. For editing, plans specify what to change and retain, while
reference audio supplies further acoustic detail.
We present \method{}, a unified framework for native First-Order
Ambisonics (FOA) generation and editing with three connected contributions.
First, a shared Transfusion backbone couples autoregressive planning with
rectified-flow rendering through executable \sceneplan{}s that specify
source content, activity, and 3D trajectories. For editing, the plan describes the full target scene, including retained
sources; contrastive features ground the requested change, while full
reference latents supply residual acoustic detail for preservation.
Second, an FOA-aware continuous VAE combines shared-decoder
omnidirectional recovery, asymmetric grouped KL, and spatial covariance
matching to preserve content and inter-channel structure.
Third, joint Transfusion on-policy self-distillation (OPSD)
refines planning on student-generated prefixes and rendering at sampling
states reached under student plans, with target-conditioned teachers
supervising both modes of the shared student.
Evaluations across speech, music, and sound show improved generation
content and spatial quality, acoustic fidelity during instruction-guided
editing, and unedited-source
preservation over the compared baselines.
Project page: \url{https://wjm-george.github.io/AMBIT/}.
\end{abstract}
\section{Introduction}
\label{sec:introduction}

Spatial audio authoring supports immersive media, virtual acoustics, and
interactive environments~\citep{zhu2025asaudio} through text-driven generation and
instruction-guided editing. Both tasks require decisions about what
each source produces, when it is active, and how it moves to be realized
in a coherent soundfield~\citep{guo2025mrsaudio}. Editing adds a further requirement: changing selected attributes while
retaining the surrounding sources.
An executable plan can state both the requested change and the sources
to keep at the semantic and structural level; reference audio then
supplies residual acoustic detail beyond these symbolic fields.
This motivates learning source decisions and multichannel synthesis
together, with plans governing what changes or stays and reference
latents supporting the acoustic preservation of retained sources.

\begin{figure}[!htbp]
    \centering
    \includegraphics[width=1.0\linewidth]{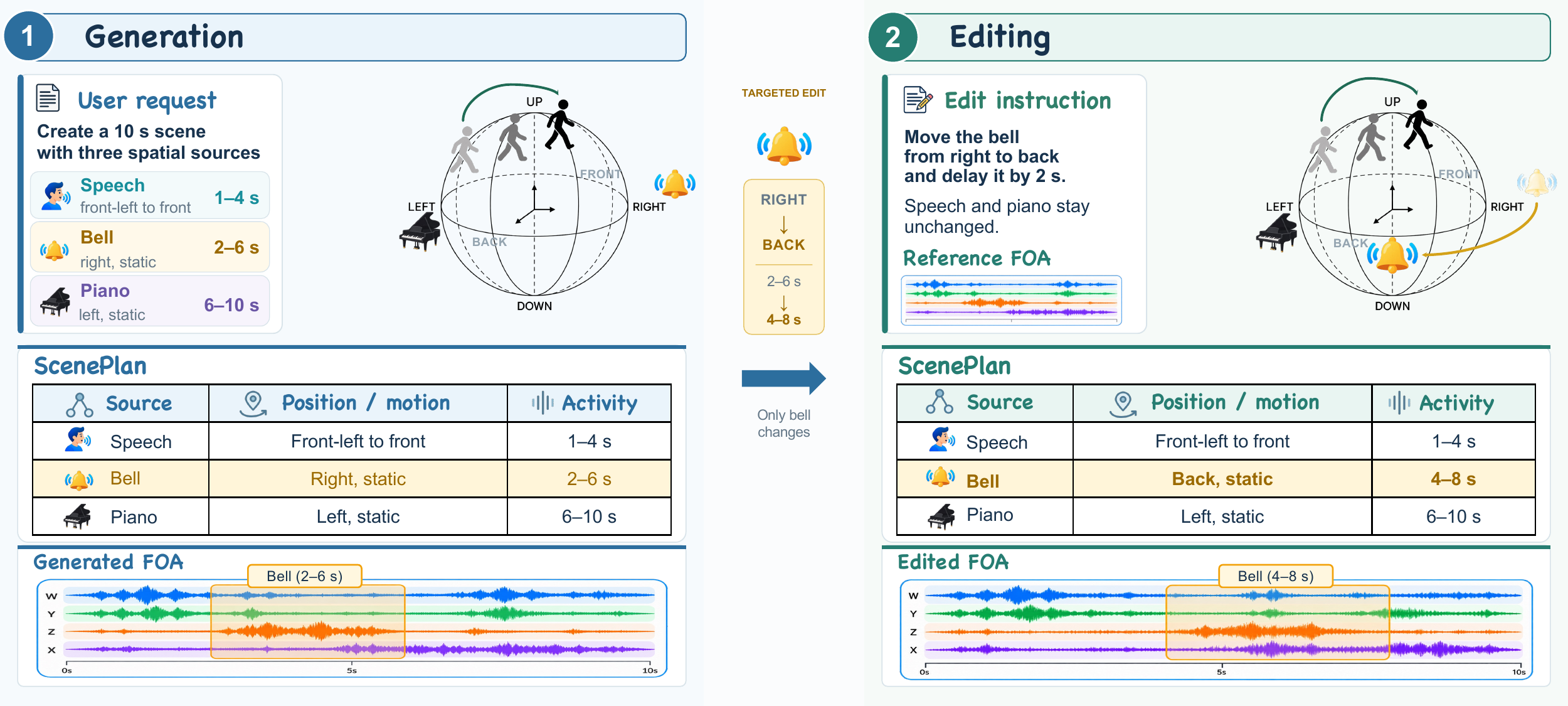}
    \caption{Overview of \method{} capabilities.
    \textbf{(1) Spatial audio generation} creates multi-source 3D scenes from text, controlling content, timing, and trajectories.
    \textbf{(2) Instruction-guided spatial and temporal editing} modifies selected sources, preserving unmodified content and spatial structure.}
    \label{fig:teaser}
\end{figure}

Recent work provides increasingly rich spatial audio controls.
ImmerseDiffusion synthesizes First-Order Ambisonics (FOA) soundscapes
from text and spatial, temporal, and environmental conditions using
latent diffusion~\citep{heydari2025immersediffusion}.
SonicMotion extends FOA generation to moving sound sources with explicit
trajectory control~\citep{templin2026generating}.
SwanWeave follows natural language instructions to edit content, timing,
and spatial attributes of existing FOA scenes~\citep{lei2026swanweave}.
Unified authoring brings these capabilities together: language intent,
source attributes, and rendered acoustics must remain aligned as scenes
are generated and modified with source-level instructions.

\noindent\textbf{(i) Grounding language in 3D source controls.}
AudioChat explores stereo planning~\citep{chen2026audiochat}; FOA authoring requires source-bound content, timing, and 3D motion, with edit targets identified from source audio.

\noindent\textbf{(ii) Preserving acoustics and spatial structure in compact latents.}
Continuous FOA autoencoders support generation, while discrete codecs provide spatial-consistency supervision~\citep{heydari2025immersediffusion,sudarsanam2025foa}.
Yet, channel-wise reconstruction fails to explicitly supervise inter-channel energy, coherence, and phase.

\noindent\textbf{(iii) Refining discrete planning and continuous rendering.}
Paired training uses reference prefixes and target plans, whereas inference
relies on student predictions. A predicted plan changes both the conditions
given to the renderer and the sampling states it encounters.
DiffusionOPSD~\citep{zhou2026policy} refines continuous generators without
discrete token supervision. Joint refinement needs token and audio targets on student prefixes
and visited sampling states.

We introduce \method{} (Figure~\ref{fig:teaser}), a unified FOA
authoring framework with three contributions.

\noindent\textbf{(i) Source-grounded Transfusion planning and rendering.}
Executable \sceneplan{}s specify source content, timing, and motion.
Their deterministically compiled controls connect autoregressive planning
to rectified-flow rendering~\citep{ma2025janusflow} within a shared backbone.
For editing, the planner predicts the complete target scene, so modified
and retained sources are both explicit in the plan; semantic and spatial
contrastive features ground the requested change, while full reference
latents provide residual acoustic detail for the preservation of unchanged sources.

\noindent\textbf{(ii) An FOA-aware continuous Variational Autoencoder.}
The latents retain acoustics and inter-channel structure. Auxiliary $W$
reconstruction reuses the decoder, and asymmetric grouped KL gives a
latent group weaker regularization for $W$ recovery. Spatial covariance
matching supervises relative energy, coherence, and phase between channels.

\noindent\textbf{(iii) Joint Transfusion On-Policy Self-Distillation.}
We adapt on-policy self-distillation~\citep{zhao2026self,jiang2026d} to both modes, with student plans driving DiT trajectories; frozen teachers use target plan memory and target
audio reference tokens to supervise student prefixes and visited states.
Token and clean-latent losses update shared and task-specific parameters,
coupling updated plans to rendering conditions and states in fresh rollouts.

For music and sound generation, \method{} achieves higher paired CLAP and lower FAD than AudioX-Turbo~\citep{tian2026audioxturbo}, while achieving the lowest speech WER and CER. Spatially, it reduces directional and binaural errors over TangoFlux~\citep{hung2026tangoflux} with spatialization.
For editing, it outperforms SwanWeave~\citep{lei2026swanweave} in acoustic fidelity and spatial accuracy.
Fixed-renderer comparisons support our planner, and VAE ablations improve
reconstruction and spatial accuracy. Joint OPSD leads continued SFT, ReFL,
and GRPO across all nine metrics.

\section{Related Work}
\label{sec:related}

\paragraph{Spatial audio generation and representation.}
General-purpose and multimodal generators broaden audio content and
conditioning~\citep{liu2024audioldm,evans2025stable,tian2026audiox,
tian2026audioxturbo,tian2025vidmuse,hung2026tangoflux,zhang2026swantale}, while timing-aware
methods introduce event-level temporal
control~\citep{xie2024picoaudio,jiang2025freeaudio,jiang2026controlaudio}.
Diff-SAGe, DualSpec, and PhysWave~\citep{kushwaha2025diff,zhao2026dualspec,yao2026physwave} explore category and location
conditioning, joint spectrogram modeling, and direction- and
distance-consistency priors,
respectively.
Video-conditioned and trajectory-aware models further extend FOA
synthesis~\citep{kim2025visage,kim2026towards,lei2026towards}.
Neural audio codecs~\citep{kumar2023high,defossez2022high,ji2025wavtokenizer}
and continuous FOA autoencoders~\citep{heydari2025immersediffusion}
provide compact representations, while FOA Tokenizer incorporates
spatial consistency supervision into discrete
compression~\citep{sudarsanam2025foa}.
\method{} combines source-level 3D control with acoustic fidelity
across native FOA generation and editing.
Our VAE supervises acoustic reconstruction and inter-channel consistency.

\paragraph{Audio editing and executable scene planning.}
AudioSpa and TAS offer text-guided mono-to-binaural spatialization~\citep{feng2025audiospa,pan2025wild}.
Diffusion editing uses noise perturbation, inversion, or instruction conditioning~\citep{meng2021sdedit,manor2024zero,jia2025audioeditor,wang2023audit,pan2026audioeditsurvey}.
MMEdit, Audio-Omni, and SmartDJ~\citep{tao2025mmedit,tian2026audio,lan2025guiding} extend language-guided editing, while SwanWeave targets native FOA editing~\citep{lei2026swanweave}.
Scripting and structured planning support compositional synthesis~\citep{liu2025wavjourney,wang2026unified,zhang2025isdrama}; AudioChat unifies stereo generation and editing~\citep{chen2026audiochat}.
Following Transfusion~\citep{zhou2025transfusion}, \method{} unifies autoregressive planning and native FOA flow rendering.
Executable, grammar-constrained \sceneplan{}s bind content, activity, and 3D paths, compiling deterministically into source-indexed semantic and frame-aligned spatial controls.
Audio--text contrastive learning aids grounding~\citep{elizalde2023clap,wu2023large,seki2026spatial,zhang2025tcsinger2}.
FOA features ground edits; reference latents support preservation.

\paragraph{On-policy refinement and self-distillation.}
Preference optimization and reinforcement learning improve audio
generation~\citep{majumder2024tango,li2026resonate}; ReFL adds
differentiable reward feedback~\citep{xu2023imagereward}.
GRPO's group-relative advantages~\citep{shao2024deepseekmath}
extend to flow and diffusion
generators~\citep{liu2026flow,xue2025dancegrpo}.
On-policy distillation supervises student-generated
sequences~\citep{agarwal2024policy}, with self-distillation for
LLM reasoning~\citep{zhao2026self} and continuous
generators~\citep{zhang2026self,zhou2026policy,jiang2026d,liu2026opsd}.
Transfusion OPSD instead jointly refines
a shared FOA editor's discrete planner and continuous renderer.
Frozen planning and rendering teachers use target plan memory
and target audio to supply token distributions
on student prefixes and clean-latent targets at flow states reached
by rendering the student's own plans.
\section{Methodology}
\label{sec:method}

\begin{figure}
    \centering
    \includegraphics[width=\linewidth]
{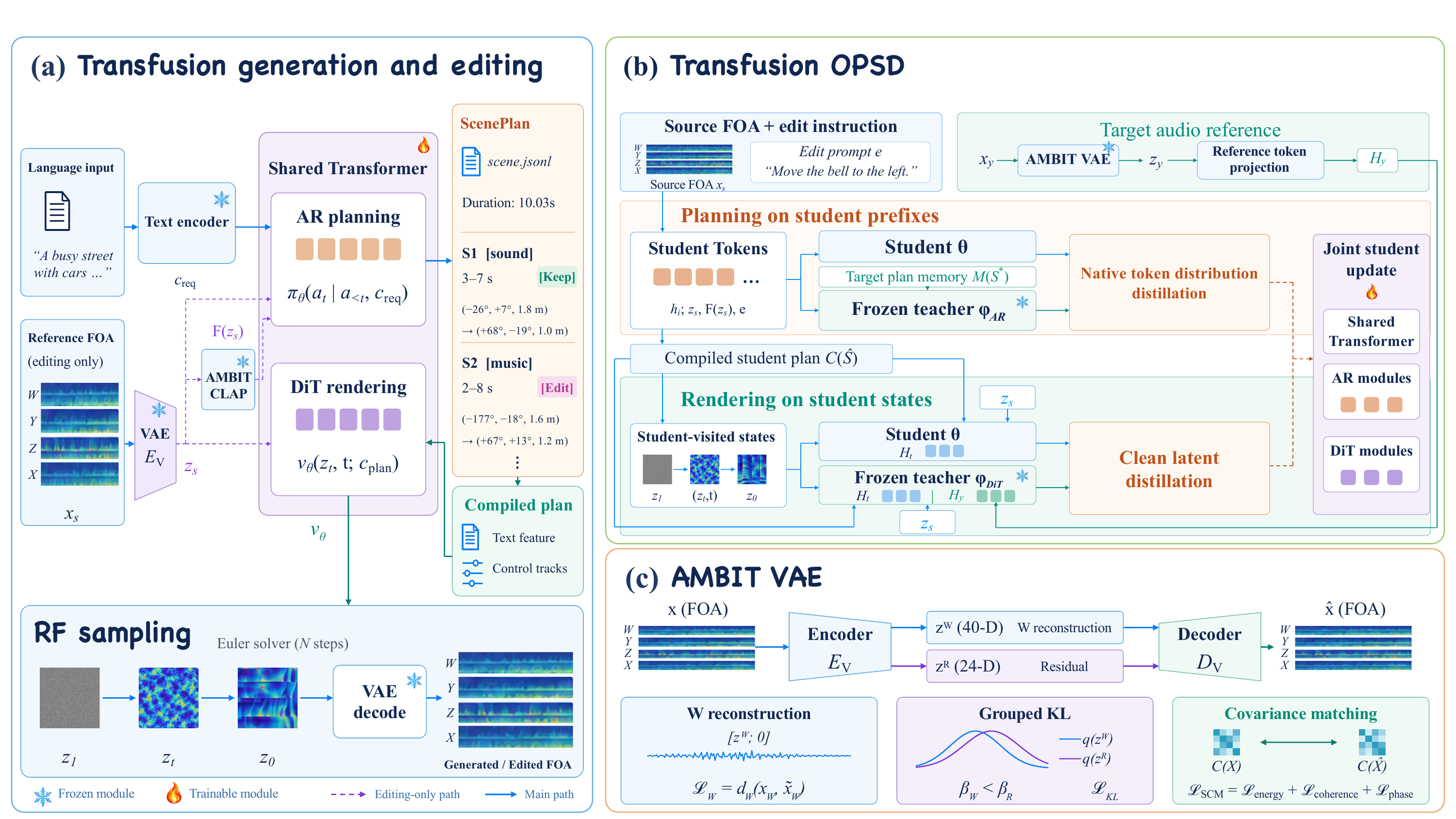}
\caption{Overview of \method{}.
\textbf{(a)} A shared Transformer predicts executable \sceneplan{}s
and renders native FOA through latent flow conditioned on compiled plans.
Editing uses source latents in both modes and \method{} CLAP features
in planning.
\textbf{(b)} OPSD distills token and clean latent predictions on student
prefixes and visited states, jointly updating the shared Transformer
and AR/DiT modules.
\textbf{(c)} The VAE combines shared-decoder auxiliary $W$ reconstruction,
grouped KL, and spatial covariance supervision.}
    \label{fig:pipeline}
\end{figure}

\paragraph{Overview.}
Let $u,e$ denote generation and editing instructions, $x_s$ a reference
First-Order Ambisonics (FOA) waveform, and $z_s=E_V(x_s)$ its latent,
with VAE encoder/decoder $E_V,D_V$.
The autoregressive (AR) planner $\pi_\theta$ predicts an executable
\sceneplan{} $\mathcal S$ token by token, specifying source-indexed
content, activity, and 3D motion.
Compilation and condition encoding produce semantic tokens and
frame-aligned spatial controls $\mathcal C(\mathcal S)$.
We adapt the Diffusion Transformer (DiT)
architecture~\citep{peebles2023scalable} for audio rendering.
Our renderer $v_\theta$ predicts audio latent velocities, which a
rectified flow sampler $\Phi[v_\theta]$ integrates before VAE decoding.
AR and DiT are two modes of a Transfusion-style Transformer,
with shared blocks and task-specific interfaces collectively
parameterized by $\theta$.
For editing, frozen \method{} CLAP features $F(z_s)$ ground instructions,
while full reference latents supply residual acoustic detail beyond
symbolic plan fields.
With $\xi=u$ for generation or $\xi=(z_s,F(z_s),e)$ for editing,
inference is
\begin{equation}
 \widehat{\mathcal S}=\pi_\theta(\xi),\qquad
 \widehat x=D_V\!\left(\Phi[v_\theta]
 \big(\epsilon;\mathcal C(\widehat{\mathcal S}),z_s\big)\right).
 \label{eq:taskgraphs}
\end{equation}
Here $\epsilon\sim\mathcal N(0,I)$ is noise and $\widehat x$ is FOA output;
generation omits $z_s,F(z_s)$ (Figure~\ref{fig:pipeline}).

\subsection{Unified Transfusion Planning and Rendering}
\label{sec:transfusion}
\label{sec:formulation}

\method{} shares a Transfusion backbone~\citep{zhou2025transfusion}
so that planning and rendering supervision shape a common representation.
An executable plan specifies the target scene; reference audio supplies
acoustic context for editing.
Mode-specific interfaces use causal attention for plan tokens and
bidirectional attention for rendering latents.

\paragraph{Executable \sceneplan{} interface.}
A \sceneplan{} $\mathcal S$ is the executable interface between
source decisions and rendering. It specifies scene duration and room
type, binding each source's type (speech, music, or sound), description,
optional transcript, active interval, and static or linear 3D trajectory
in azimuth, elevation, and distance. For editing, the planner predicts
the complete target scene, including both modified and retained source attributes.

\paragraph{\sceneplan{} compilation.}
The compiler deterministically forms source text clauses and aligns
activity and pose to latent frames (Appendix~\ref{app:compiler}).
Condition encoding uses a frozen text backbone with a learned projection
and source role embeddings to produce cross-attention tokens
$c_{\mathrm{attn}}$. Learned event embeddings encode interval-overlap
activity; a trajectory MLP encodes poses at frame midpoints.
The spatial embeddings form $c\in\mathbb R^{256\times T}$ over $T$
latent frames. Together they yield $\mathcal C(\mathcal S)=
(c_{\mathrm{attn}}(\mathcal S),c(\mathcal S))$
with semantic tokens and frame-aligned source controls.

\paragraph{\method{} CLAP for source understanding.}
Following CLAP~\citep{elizalde2023clap,wu2023large}, we learn
separate semantic and scene alignment heads from frozen
\method{} VAE latents:
\begin{equation}
  \mathcal L_{\mathrm{align}}
  =\mathcal L_{\mathrm{sem}}
  +\lambda_{\mathrm{scene}}\mathcal L_{\mathrm{scene}},
  \label{eq:foa-clap-alignment}
\end{equation}
Both terms use symmetric audio--text contrastive losses with multiple
positives. Semantic alignment matches content and source counts;
scene alignment, weighted by $\lambda_{\mathrm{scene}}$, also matches
timing, spatial attributes, and room properties.
Content-preserving spatial edits remain semantic positives but become
scene negatives. The frozen encoder supplies global and temporal
features $F(z_s)$ for semantic and spatial grounding in the editing
planner (Appendix~\ref{app:foa-clap}).

\paragraph{Autoregressive planning.}
A grammar constrained codec serializes \sceneplan{}s into tokens.
We train the planner with next token cross entropy
$\mathcal L_{\mathrm{AR}}$ on reference prefixes and task inputs.
Planning heads predict source inventory, copy literal spans, and select
qualitative controls for deterministic completion
(Appendix~\ref{app:planner}).
Editing uses a bidirectional source latent prefix with \method{} CLAP
features and instruction cross attention.
A source decoder predicts edit targets, operations, and updated source
attributes; its memory guides AR decoding through gated residual attention.

\paragraph{Flow rendering.}
A compiled \sceneplan{} specifies scene structure; full reference
latents $z_s$ provide acoustic details of retained sources for editing.
For target waveform $x_y$, let $z_0=z_y=E_V(x_y)$ be its clean latent.
We sample flow time $t\in[0,1]$ from $p(t)$, construct
$z_t=(1-t)z_0+t\epsilon$, and minimize the base rectified flow
objective~\citep{liu2022flow}:
\begin{equation}
  \mathcal L_{\mathrm{RF}}
  =\mathbb E\!\left[
  \frac{1}{\|m\|_1}
  \left\|m\odot\left(
  v_\theta\!\left([z_t;c(\mathcal S);z_s],t
  \mid c_{\mathrm{attn}}(\mathcal S)\right)
  -(\epsilon-z_0)\right)\right\|_2^2\right].
  \label{eq:rf}
\end{equation}
Brackets concatenate channels; $\odot$ multiplies elementwise.
The binary mask $m$ selects valid latent coordinates across all channels,
excluding padded frames; $\|m\|_1$ counts all valid latent coordinates.
For generation, the reference-latent channels are set to zero.
Sampling integrates from noise at $t=1$ to the output latent at $t=0$.
Generation training uses normalized speech activity weighting and an
auxiliary latent temporal difference loss for sound only scenes;
Appendices~\ref{app:generation} and~\ref{app:training} specify these
objectives and the flow time distributions used for both training tasks.

\paragraph{Joint planning and rendering.}
Paired editing jointly supervises plan tokens, audio velocities,
recognition, edit binding, and preservation. Structured supervision
reuses one source assignment for retained and modified attributes:
\begin{equation}
  \mathcal L_{\mathrm{joint}}
  =\mathcal L_{\mathrm{AR}}+\mathcal L_{\mathrm{RF}}
  +\lambda\mathcal L_{\mathrm{struct}},
  \label{eq:edit-joint}
\end{equation}
where $\lambda$ weights the structured source supervision
$\mathcal L_{\mathrm{struct}}$ (Appendix~\ref{app:editing}).
AR and DiT losses update shared blocks and mode interfaces.
Paired rendering uses ground truth target plans; inference
uses student predictions.
Text backbones, \method{} CLAP, and the VAE remain frozen;
student condition projections and embeddings remain trainable.

\subsection{\method{} Variational Autoencoder}
\label{sec:foa-vae}

We adapt the continuous audio VAE of Stable Audio
Open~\citep{evans2025stable} to encode four-channel waveforms
$x$ as $z=E_V(x)$ and reconstruct them through $D_V(z)$.
Extending the input and output to four channels alone does not
prioritize omnidirectional $W$ recovery or provide explicit supervision
of FOA inter-channel relationships.
We therefore combine auxiliary $W$ reconstruction and asymmetric
grouped KL to support acoustic recovery, with spatial covariance
matching (SCM) to supervise relative channel energy, coherence, and
phase during reconstruction.

\paragraph{Omnidirectional recovery and capacity allocation.}
The omnidirectional $W$ channel provides a scene-wide acoustic
reconstruction target.
We partition $z=[z^W;z^R]$, where $z^R$ contains the remaining
latent channels.
The full latent reconstructs all four channels.
An auxiliary pass masks $z^R$ and supervises only $W$ through
the same decoder; we combine this auxiliary reconstruction objective with grouped KL regularization:
\begin{equation}
 \widetilde x_W=\big[D_V([z^W;\mathbf 0])\big]_W,\qquad
 \mathcal L_{\mathrm{KL}}=\beta_W K_W+\beta_R K_R,
 \quad \beta_W<\beta_R.
 \label{eq:vae-w}
\end{equation}
$\mathcal L_W$ combines multiresolution
short-time Fourier transform (STFT) and SI-SDR
losses~\citep{le2019sdr}; $K_W,K_R$ are groupwise KL penalties.
The $W$ task assigns $z^W$ a recovery role under weaker KL.
Decoder reuse adds no parameters; the partition encourages specialization
without enforcing exact content--space disentanglement.

\paragraph{Spatial covariance matching.}
Auxiliary $W$ recovery constrains omnidirectional acoustics; SCM adds
supervision of inter-channel structure. It compares local STFT covariance through energy
fractions $p_i$, coherence magnitudes $\gamma_{ij}$, and stabilized
inter-channel phase factors $U_{ij}$ between reference and reconstructed audio:
\begin{equation}
 \mathcal L_{\mathrm{SCM}}
 =\Big\langle\sum_i|\widehat p_i-p_i|\Big\rangle_w
 +\Big\langle\sum_{i<j}
   |\widehat\gamma_{ij}-\gamma_{ij}|\Big\rangle_w
 +\Big\langle 1-\operatorname{Re}
   (\widehat U_{ij}\overline{U_{ij}})\Big\rangle_v.
 \label{eq:scm-main}
\end{equation}
Hats denote reconstructed statistics; brackets denote weighted averages.
The phase component retains magnitude dependence from stabilization.
The full VAE training objective is
\begin{equation}
 \mathcal L_{\mathrm{VAE}}
 =\mathcal L_{\mathrm{base}}+\mathcal L_W+\mathcal L_{\mathrm{KL}}
 +\lambda_{\mathrm{SCM}}\mathcal L_{\mathrm{SCM}}.
 \label{eq:vae-total}
\end{equation}
Here $\mathcal L_{\mathrm{base}}$ contains spectral, waveform, and
adversarial terms, and $\lambda_{\mathrm{SCM}}$ weights spatial
supervision. Appendix~\ref{app:vae} gives covariance statistics, stabilization,
reference weights, and reductions across channels, time, and resolutions.

\subsection{Joint Transfusion On-Policy Self-Distillation}
\label{sec:joint-refinement}
Paired training uses reference prefixes and ground-truth plans,
whereas inference uses the student's own decisions.
These decisions determine both the conditions given to DiT and the
sampling states reached under them.
We adapt on-policy self-distillation~\citep{zhao2026self,jiang2026d}
to this coupled process: student plans drive DiT trajectories, and
both modes receive supervision on the contexts the student encounters.
We prepare both teachers from a common supervised model: the planner
uses target plan memory, and the renderer uses audio reference tokens.
Teachers are then frozen; the editing student uses source audio and
instructions during rollouts and inference.

\paragraph{Planning on student prefixes.}
The student and planning teacher $\phi_{\mathrm{AR}}$ share editing
conditions $c=(z_s,F(z_s),e)$ and a student generated prefix $h_i$.
The teacher additionally reads target plan memory $M(\mathcal S^\star)$.
For non-forced decisions $\mathcal I$ and student distribution
$p_{\theta,i}=p_\theta(\cdot\mid h_i,c)$, we use
\begin{equation}
  q_i=\operatorname{sg}\!\left[
  p_{\phi_{\mathrm{AR}}}(\cdot\mid h_i,c,M(\mathcal S^\star))\right],
  \qquad
  \ell_{\mathrm{AR}}=\sum_{i\in\mathcal I}\eta_i
  D_{\mathrm{KL}}\!\left(q_i\,\|\,p_{\theta,i}\right).
  \label{eq:opsd-local-teacher}
\end{equation}
Both distributions are normalized over the same tokens allowed
by the grammar at $h_i$.
The stop-gradient operator $\operatorname{sg}$ treats $q_i$ as a
fixed teacher target, so the KL loss updates only the student.
Detached weights $\eta_i$ balance decision families and exclude
positions where aligned target facts are illegal or conflict with
teacher predictions.
A reference loss $\ell_{\mathrm{ref}}$ anchors selected plan fields
to predictions of the frozen initial supervised model
on $h_i$ under ordinary conditions.

\paragraph{Rendering on student states.}
Compiling the AR student's plan $\widehat{\mathcal S}$ provides conditions
for DiT trajectories alongside source latent $z_s$.
We collect visited states $(z_j,t_j)$, with queries $\mathcal Q$
stratified across sampling times.
The teacher $\phi_{\mathrm{DiT}}$ receives the same states,
source latent, and plan, plus temporally aligned reference tokens
$H_y=P_{\phi_{\mathrm{DiT}}}(z_y)$ projected from clean target
audio latents as additional attention context.
Matching the models' clean latent predictions gives
\begin{equation}
  \ell_{\mathrm{DiT}}=\sum_{j\in\mathcal Q}
  \frac{\omega_j t_j^2}{\|m\|_1}
  \Big\|m\odot\Big(
  v_\theta(z_j,t_j;\widehat{\mathcal S},z_s)
  -\operatorname{sg}\!\left[
  v_{\phi_{\mathrm{DiT}}}(z_j,t_j;\widehat{\mathcal S},z_s,H_y)
  \right]\Big)\Big\|_2^2.
  \label{eq:opsd-terminal-flow}
\end{equation}
Masking and normalization follow Equation~\ref{eq:rf}.
The factor $t_j^2$ follows from $\widehat z_0=z_j-t_jv$;
$\omega_j$ averages noise trajectories and time strata.
Sampled states, plans, and teacher outputs are detached during
student optimization.

\paragraph{Joint refinement through shared parameters.}
For training records $(x_s,e,\mathcal S^\star,x_y)\sim\mathcal D$,
\begin{equation}
 \mathcal L_{\mathrm{OPSD}}
 =\mathbb E_{\mathcal D}\!\left[
 \lambda_{\mathrm{AR}}\ell_{\mathrm{AR}}
 +\lambda_{\mathrm{DiT}}\ell_{\mathrm{DiT}}
 +\lambda_{\mathrm{ref}}\ell_{\mathrm{ref}}\right].
 \label{eq:opsd-objective}
\end{equation}
The losses jointly update the shared Transformer and active task modules.
Rendering gradients shape the representation used for later planning,
while updated plans change subsequent rendering conditions and states.
This coupling occurs through shared updates and fresh rollouts;
sampled plans and states remain detached.
Only student predictions and trainable conditions receive gradients;
teachers, reference models, and pretrained backbones remain frozen.
Appendix~\ref{app:opsd} gives implementation details and a finite-step
teacher-tracking analysis under student plans.

\providecommand{\ambitmetriccell}[1]{%
  \begin{tabular}[c]{@{}c@{}}#1\end{tabular}%
}

\begingroup
\setlength{\intextsep}{6pt plus 1pt minus 1pt}
\setlength{\textfloatsep}{7pt plus 1pt minus 1pt}
\setlength{\floatsep}{6pt plus 1pt minus 1pt}
\setlength{\abovecaptionskip}{3pt}
\setlength{\belowcaptionskip}{2pt}
\renewcommand{\topfraction}{0.80}
\renewcommand{\bottomfraction}{0.55}
\renewcommand{\textfraction}{0.15}
\renewcommand{\floatpagefraction}{0.95}
\setcounter{topnumber}{2}
\setcounter{bottomnumber}{1}
\setcounter{totalnumber}{2}

\section{Experiments}
\label{sec:experiments}
\label{sec:results}

\subsection{Experimental Setup}
\label{sec:experimental-setup}

\noindent\textbf{Datasets.}
Training uses 1.6M synthetic and recorded FOA examples and 1M editing
pairs, covering speech, music, and environmental sounds.
The \method{} VAE is pretrained on 1M FOA clips.
Evaluation uses 8,000 generation requests, 5,000 editing pairs,
and 2,000 VAE test clips.
Appendices~\ref{app:data} and~\ref{app:exp-setup} specify the data
and evaluation protocols.

\paragraph{Model and training.}
\method{} uses 15 shared Transfusion blocks
(width 1,024, 16 attention heads) for native FOA generation and editing.
The shared backbone, native AR/DiT interfaces, and structured heads
total 332.6M trainable parameters. Joint supervised training uses
token and rectified flow objectives with AdamW at $10^{-4}$ for
adapters and heads and $2\times10^{-6}$ for shared blocks.
Paired rendering uses ground-truth plans; inference uses AR predictions.
The Qwen3.5-0.8B text backbone \citep{qwen2026qwen35},
\method{} CLAP, and \method{} VAE remain frozen.
OPSD uses AdamW with a global batch of 64 requests and loss weights
$\lambda_{\mathrm{AR}}=0.1$, $\lambda_{\mathrm{DiT}}=1$, and
$\lambda_{\mathrm{ref}}=0.1$, with prepared teachers frozen.
Appendices~\ref{app:implementation}, \ref{app:training},
and~\ref{app:opsd} specify the architecture, supervised training,
and post-training configurations.

\begin{table}[!htb]
\centering
\small
\setlength{\tabcolsep}{3.5pt}
\renewcommand{\arraystretch}{1.04}
\caption{FOA reconstruction and cumulative ablation.
The upper block compares codecs on native scenes; each ``+'' below
adds to the previous row, with grouped KL replacing uniform KL.
Ablations use matched updates from scratch and 24\,kHz evaluation.
Bold marks the best per block.}
\label{tab:vae-all}
\label{tab:vae-objective-ablation}
\begin{tabularx}{\linewidth}{c*{4}{>{\centering\arraybackslash}X}}
\toprule
\ambitmetriccell{System/configuration}
& \ambitmetriccell{$W$ SI-SDR\\(dB) $\uparrow$}
& \ambitmetriccell{Direction\\($^\circ$) $\downarrow$}
& \ambitmetriccell{Intensity\\coherence error $\downarrow$}
& \ambitmetriccell{$W$ paired\\CLAP $\uparrow$} \\
\midrule
DAC & 5.57 & 5.13 & 0.08 & 0.88 \\
EnCodec & 0.41 & 12.25 & 0.15 & 0.85 \\
WavTokenizer & $-19.41$ & 56.66 & 0.35 & 0.63 \\
\rowcolor{blue!10}
\method{} VAE
& \textbf{6.39} & \textbf{4.96}
& \textbf{0.04} & \textbf{0.90} \\
\midrule
Phase (base) & 5.63 & 6.51 & 0.06 & 0.86 \\
+ SCM & 6.08 & 5.14 & 0.05 & 0.88 \\
+ Cepstral & 6.15 & 5.09 & 0.06 & 0.87 \\
+ SI-SDR & 6.25 & 5.08 & 0.05 & 0.87 \\
\rowcolor{blue!10}
+ Grouped KL + $\mathcal L_W$ (full)
& \textbf{6.39} & \textbf{4.96}
& \textbf{0.04} & \textbf{0.90} \\
\bottomrule
\end{tabularx}
\end{table}

\noindent\textbf{Evaluation metrics.}
We evaluate generation and editing with the same \method{} checkpoint. For spatial audio, we follow SwanSphere~\citep{lei2026towards}
in reporting FD-PaSST and KL-PaSST, supplemented by paired audio
CLAP, FAD, KL-PANNs, and FD-OpenL3.
Speech evaluation uses Distil-Whisper
WER/CER~\citep{gandhi2023distil} and UTMOS~\citep{saeki2022utmos}.
For editing, we report FD, FAD, KL, and LSD following
SwanWeave~\citep{lei2026swanweave}, with CLAP and PANNs features
for FD, and the paired CLAP score~\citep{wu2023large}.
Both tasks report direction errors on eligible isolated intervals:
spherical and azimuth MAE, and angular trajectory RMSE.
Complementary GCC, CRW~\citep{chen2022sound},
FSAD~\citep{sun2025both}, and wCS from
PSELDNets~\citep{lei2026towards,hu2025pseldnets}, together with
scene direction and unedited source
preservation, are detailed in Appendix~\ref{app:exp-setup}.
Codec evaluation reports reconstruction quality, intensity-vector
direction error, and framewise intensity-coherence error
in Eq.~\ref{eq:intensity-coherence-error}.
Appendix~\ref{app:user-study} complements these metrics with listener
ratings of audio quality, instruction adherence, and preservation.

\subsection{\method{} VAE Evaluation}
\label{sec:vae-evaluation}
\label{sec:vae-ablation}
\label{sec:analysis}

FOA reconstruction must preserve acoustic content and spatial structure.
We compare codecs and train cumulative objective variants from scratch
for matched updates (Table~\ref{tab:vae-all}).
\method{} outperforms DAC~\citep{kumar2023high},
EnCodec~\citep{defossez2022high}, and
WavTokenizer~\citep{ji2025wavtokenizer} across all four metrics.

SCM reduces direction error from $6.51^\circ$ to $5.14^\circ$,
while increasing $W$ SI-SDR from 5.63 to 6.08\,dB.
Subsequent cepstral and SI-SDR terms raise $W$ SI-SDR to 6.25\,dB,
with only a small further directional gain ($5.08^\circ$).
This pattern highlights the value of explicitly supervising interchannel
relationships alongside acoustic reconstruction.
Grouped KL with auxiliary $W$ recovery then improves all four metrics,
supporting stronger omnidirectional reconstruction together with spatial
fidelity (Appendix~\ref{app:vae-protocol}).

\subsection{Generation and Editing Quality}
\label{sec:content-quality}
\label{sec:results-gen}

We examine whether executable plans support content fidelity across
music, sound, and speech through comparisons with domain baselines
(Table~\ref{tab:quality}).
\method{} leads paired CLAP, classifier KL, and FAD in music and sound,
while AudioX-Turbo retains lower FD-PaSST/FD-OpenL3.
For speech, \method{} achieves lower WER/CER but lower UTMOS than Qwen3-TTS.
These results align with source descriptions and literal
transcripts specifying event content and spoken words: semantic and
lexical fidelity improve, with varying perceptual and distributional rankings.

\begin{table}[!htb]
\centering
\small
\setlength{\tabcolsep}{1.8pt}
\renewcommand{\arraystretch}{1.08}
\caption{Generation content quality.
Bold and underlined values indicate the best and second-best results
within each evaluated domain; displayed ties receive the same marking.}
\label{tab:quality}
\begin{tabularx}{\linewidth}{cc*{6}{>{\centering\arraybackslash}X}}
\toprule
Domain & System
& \ambitmetriccell{Paired\\CLAP $\uparrow$}
& FAD $\downarrow$
& \ambitmetriccell{KL\\PANNs $\downarrow$}
& \ambitmetriccell{KL\\PaSST $\downarrow$}
& \ambitmetriccell{FD\\PaSST $\downarrow$}
& \ambitmetriccell{FD\\OpenL3 $\downarrow$} \\
\midrule
\multirow{6}{*}{Music} & AudioX-Turbo~\citep{tian2026audioxturbo} & \underline{0.36} & 1.07 & 0.68 & 1.58 & \textbf{234.65} & \textbf{37.40} \\
 & AudioX~\citep{tian2026audiox} & 0.35 & 1.30 & 0.74 & 1.56 & \underline{243.68} & 46.67 \\
 & Stable Audio Open~\citep{evans2025stable} & 0.18 & 2.08 & 0.90 & 1.81 & 410.64 & 68.72 \\
 & TangoFlux~\citep{hung2026tangoflux} & \underline{0.36} & \underline{1.01} & 0.59 & 1.43 & 319.37 & 92.09 \\
 & BEWO~\citep{sun2025both} & 0.26 & 2.05 & \underline{0.56} & \underline{1.40} & 566.57 & 57.30 \\
\rowcolor{blue!10}
 & \method{} & \textbf{0.46} & \textbf{0.71} & \textbf{0.51} & \textbf{0.95} & 261.45 & \underline{44.59} \\
\midrule
\multirow{6}{*}{Sound} & AudioX-Turbo~\citep{tian2026audioxturbo} & 0.34 & \underline{0.79} & 0.63 & 2.36 & \textbf{213.60} & \textbf{40.56} \\
 & AudioX~\citep{tian2026audiox} & 0.32 & 1.08 & 0.68 & 2.35 & \underline{228.62} & 52.17 \\
 & Stable Audio Open~\citep{evans2025stable} & 0.17 & 1.58 & 0.79 & 2.70 & 334.40 & 61.81 \\
 & TangoFlux~\citep{hung2026tangoflux} & \underline{0.36} & 0.99 & \underline{0.54} & \underline{2.18} & 243.97 & 85.85 \\
 & BEWO~\citep{sun2025both} & 0.21 & 2.40 & 0.55 & 2.45 & 517.14 & 66.27 \\
\rowcolor{blue!10}
 & \method{} & \textbf{0.41} & \textbf{0.75} & \textbf{0.53} & \textbf{1.78} & 235.49 & \underline{49.99} \\
\midrule
Domain & System
& \multicolumn{2}{c}{WER $\downarrow$}
& \multicolumn{2}{c}{CER $\downarrow$}
& \multicolumn{2}{c}{UTMOS $\uparrow$} \\
\midrule
\multirow{4}{*}{Speech} 
 & UniAudio2~\citep{yang2026uniaudio} & \multicolumn{2}{c}{\underline{0.08}} & \multicolumn{2}{c}{\underline{0.06}} & \multicolumn{2}{c}{\underline{4.11}} \\
 & Qwen3-TTS 1.7B~\citep{hu2026qwen3} & \multicolumn{2}{c}{0.09} & \multicolumn{2}{c}{0.08} & \multicolumn{2}{c}{\textbf{4.18}} \\
 & Parler-TTS Large~\citep{lyth2024natural} & \multicolumn{2}{c}{0.11} & \multicolumn{2}{c}{\underline{0.06}} & \multicolumn{2}{c}{3.40} \\
\rowcolor{blue!10}
 & \method{} & \multicolumn{2}{c}{\textbf{0.07}} & \multicolumn{2}{c}{\textbf{0.05}} & \multicolumn{2}{c}{3.52} \\
\bottomrule
\end{tabularx}
\end{table}

\label{sec:results-edit}
Editing requires carrying out an instruction while retaining the
source's acoustic context.
We compare editing baselines (Table~\ref{tab:editing-quality}).
\method{} improves all six metrics over SwanWeave, the strongest baseline.
Higher paired CLAP and lower LSD extend the gains from semantic agreement
to spectral reconstruction, alongside improved distributional scores.
Table~\ref{tab:spatial-editing} evaluates how well the output preserves
direction, waveform, and level of retained sources.

\begin{table}[!htb]
\centering
\small
\setlength{\tabcolsep}{2pt}
\renewcommand{\arraystretch}{1.08}
\caption{Editing content quality and spectral reconstruction. The KL here is KL-PANNs.
Bold and underlined values indicate the best and second-best results.
Spatial metrics are reported in Table~\ref{tab:spatial-editing}.}
\label{tab:editing-quality}
\begin{tabularx}{\linewidth}{c
>{\hsize=1.48\hsize\linewidth=\hsize\centering\arraybackslash}X
>{\hsize=1.23\hsize\linewidth=\hsize\centering\arraybackslash}X
>{\hsize=0.79\hsize\linewidth=\hsize\centering\arraybackslash}X
>{\hsize=1.25\hsize\linewidth=\hsize\centering\arraybackslash}X
>{\hsize=0.57\hsize\linewidth=\hsize\centering\arraybackslash}X
>{\hsize=0.68\hsize\linewidth=\hsize\centering\arraybackslash}X
}
\toprule
System
& \ambitmetriccell{Paired\\CLAP $\uparrow$}
& \ambitmetriccell{FD\\CLAP $\downarrow$}
& \ambitmetriccell{FAD $\downarrow$}
& \ambitmetriccell{FD\\PANNs $\downarrow$}
& \ambitmetriccell{KL $\downarrow$}
& \ambitmetriccell{LSD $\downarrow$} \\
\midrule
Audio-Omni~\citep{tian2026audio} & 0.46 & 0.23 & 0.91 & 23.75 & 2.04 & 2.09 \\
MMEdit~\citep{tao2025mmedit} & 0.48 & 0.31 & 1.15 & 22.73 & 1.37 & 2.45 \\
SwanWeave~\citep{lei2026swanweave} & \underline{0.68} & \underline{0.15} & \underline{0.57} & \underline{11.40} & \underline{0.92} & \underline{1.65} \\
SmartDJ~\citep{lan2025guiding} & 0.63 & 0.17 & 0.70 & 19.30 & 1.16 & 1.89 \\
ZETA~\citep{manor2024zero} & 0.41 & 0.38 & 1.82 & 33.56 & 1.63 & 2.86 \\
AudioEditor--BEWO~\citep{jia2025audioeditor} & 0.44 & 0.36 & 1.87 & 25.12 & 1.39 & 2.43 \\
SDEdit~\citep{meng2021sdedit} & 0.32 & 0.48 & 2.13 & 41.45 & 1.95 & 3.36 \\
\midrule
\rowcolor{blue!10}
\method{} & \textbf{0.84} & \textbf{0.05} & \textbf{0.24} & \textbf{6.01} & \textbf{0.46} & \textbf{1.27} \\
\bottomrule
\end{tabularx}
\end{table}

\subsection{Spatial Audio Evaluation}
\label{sec:spatial-evaluation}

Spatial control depends on translating plan attributes into audio.
We compare scene and object metrics against cascaded baselines.
BEWO + AS and TangoFlux + AS use audio
spatialization (AS) with Qwen3.5-4B \sceneplan{}s and
Pyroomacoustics~\citep{scheibler2018pyroomacoustics};
Qwen3.5-4B + DiT uses our renderer (Appendix~\ref{app:spatial-cascades}).
Object metrics use intervals with one annotated active source
(Appendix~\ref{app:metrics}); planner validity appears in
Appendix~\ref{app:ar-evaluation}.
Qwen3.5-4B + DiT improves GCC, CRW, and wCS over AS pipelines
but has higher scene and object direction errors;
binaural gains thus need not imply accurate source placement.
\method{} leads both groups (Table~\ref{tab:spatial-generation}).
With the same renderer, gains over Qwen3.5-4B + DiT support the learned
planner's spatial controls at both scene and object levels.

\begin{table}[!htb]
\centering
\small
\setlength{\tabcolsep}{2.5pt}
\renewcommand{\arraystretch}{1.00}
\caption{Spatial generation at scene and object levels.
Scene direction errors use mixtures; object metrics use isolated
source intervals. Angular errors are in radians. Bold marks the best score.}
\label{tab:spatial-generation}
\begin{tabularx}{\linewidth}{c*{7}{>{\centering\arraybackslash}X}}
\toprule
& \multicolumn{7}{c}{Scene spatial quality} \\
\cmidrule(lr){2-8}
System
& GCC $\downarrow$ & CRW $\downarrow$ & FSAD $\downarrow$
& DOA $\downarrow$
& \ambitmetriccell{Azimuth\\MAE $\downarrow$}
& \ambitmetriccell{Ang. traj.\\RMSE $\downarrow$}
& wCS $\uparrow$ \\
\midrule
BEWO + AS
& 34.71
& 73.22
& 0.03
& 1.04
& 1.00
& 1.13
& 0.11 \\
TangoFlux + AS
& 36.42
& 67.88
& 0.04
& 1.03
& 0.98
& 1.12
& 0.09 \\
Qwen3.5-4B + DiT
& 31.07
& 64.36
& 0.03
& 1.07
& 1.02
& 1.18
& 0.27 \\
\midrule
\rowcolor{blue!10}
\method{}
& \textbf{12.59}
& \textbf{8.75}
& \textbf{0.02}
& \textbf{0.51}
& \textbf{0.40}
& \textbf{0.63}
& \textbf{0.56} \\
\bottomrule
\end{tabularx}

\vspace{3pt}

\begin{tabularx}{\linewidth}{c*{3}{>{\centering\arraybackslash}X}}
\toprule
& \multicolumn{3}{c}{Object direction} \\
\cmidrule(lr){2-4}
System
& \ambitmetriccell{Spherical\\MAE $\downarrow$}
& \ambitmetriccell{Azimuth\\MAE $\downarrow$}
& \ambitmetriccell{Angular trajectory\\RMSE $\downarrow$} \\
\midrule
BEWO + AS
& 1.001
& 0.949
& 1.066 \\
TangoFlux + AS
& 0.994
& 0.942
& 1.055 \\
Qwen3.5-4B + DiT
& 1.030
& 0.973
& 1.113 \\
\midrule
\rowcolor{blue!10}
\method{}
& \textbf{0.531}
& \textbf{0.415}
& \textbf{0.629} \\
\bottomrule
\end{tabularx}
\end{table}

Preserving unedited sources is a separate requirement from matching
target directions.
We compare scene quality, object directions, and preservation against
SwanWeave and Qwen3-Omni + DiT~\citep{xu2025qwen3}
(Table~\ref{tab:spatial-editing}).
The latter uses our renderer and improves scene and object metrics
and $W$ NMSE over SwanWeave, but increases unchanged-source
direction and level errors.
This motivates measuring preservation against original source audio
alongside reconstruction against the target scene.
\method{} leads all three metric groups, improving target direction
accuracy alongside retention of the original direction, waveform,
and level on eligible unchanged-source intervals.
These results align with plans specifying requested changes and retained
sources, with full reference latents supplying
residual acoustic detail for preservation.
Listener ratings also favor edit success and preservation
(Appendix~\ref{app:user-study}).

\begin{table}[!htb]
\centering
\small
\setlength{\tabcolsep}{2.2pt}
\renewcommand{\arraystretch}{1.00}
\caption{Spatial editing: scene metrics (upper), preservation and object
metrics on isolated intervals (lower).
Angular errors are in radians; level changes in dB.
Bold marks the best score.}
\label{tab:spatial-editing}
\begin{tabularx}{\linewidth}{c*{7}{>{\centering\arraybackslash}X}}
\toprule
& \multicolumn{7}{c}{Scene spatial quality} \\
\cmidrule(lr){2-8}
\ambitmetriccell{System}
& \ambitmetriccell{GCC $\downarrow$}
& \ambitmetriccell{CRW $\downarrow$}
& \ambitmetriccell{FSAD $\downarrow$}
& \ambitmetriccell{DOA $\downarrow$}
& \ambitmetriccell{Az.\\MAE $\downarrow$}
& \ambitmetriccell{Ang. traj.\\RMSE $\downarrow$}
& \ambitmetriccell{wCS $\uparrow$} \\
\midrule
SwanWeave
& 30.77 & 41.11 & 0.04 & 1.11 & 1.06 & 1.25 & 0.27 \\
Qwen3-Omni + DiT
& 20.45 & 26.82 & 0.02 & 0.86 & 0.85 & 1.01 & 0.36 \\
\midrule
\rowcolor{blue!10}
\method{}
& \textbf{14.81} & \textbf{11.95} & \textbf{0.01}
& \textbf{0.39} & \textbf{0.36} & \textbf{0.52} & \textbf{0.71} \\
\bottomrule
\end{tabularx}

\vspace{3pt}

\begin{tabularx}{\linewidth}{c*{6}{>{\centering\arraybackslash}X}}
\toprule
& \multicolumn{3}{c}{Unedited source preservation}
& \multicolumn{3}{c}{Object direction} \\
\cmidrule(lr){2-4}\cmidrule(lr){5-7}
\ambitmetriccell{System}
& \ambitmetriccell{Direction\\MAE $\downarrow$}
& \ambitmetriccell{$W$\\NMSE $\downarrow$}
& \ambitmetriccell{Abs. level\\change $\downarrow$}
& \ambitmetriccell{Spherical\\MAE $\downarrow$}
& \ambitmetriccell{Azimuth\\MAE $\downarrow$}
& \ambitmetriccell{Angular traj.\\RMSE $\downarrow$} \\
\midrule
SwanWeave
& 0.90 & 1.49 & 8.40 & 1.053 & 1.016 & 1.157 \\
Qwen3-Omni + DiT
& 1.42 & 0.58 & 20.94 & 0.734 & 0.711 & 0.797 \\
\midrule
\rowcolor{blue!10}
\method{}
& \textbf{0.17} & \textbf{0.44} & \textbf{1.65}
& \textbf{0.200} & \textbf{0.154} & \textbf{0.270} \\
\bottomrule
\end{tabularx}
\end{table}

\subsection{Post-training with OPSD}
\label{sec:opsd-evaluation}

Post-training should improve how scene decisions are realized in audio.
We compare OPSD configurations with continued SFT,
our ReFL adaptation~\citep{xu2023imagereward}, and
two branch GRPO~\citep{shao2024deepseekmath,liu2026flow} under
identical evaluation settings (Appendix~\ref{app:post-training-controls}).
Joint OPSD leads all reported metrics (Table~\ref{tab:opsd-editing}).
DiT OPSD improves every metric over SFT, whereas AR OPSD improves
only LSD and FSAD.
The gains across content, reconstruction, and binaural metrics support
refining planning and rendering together within the shared editing model.

A further question is whether on-policy planning distillation adds to rendering refinement under student plans. We contrast joint OPSD with the two supervision
variants. AR OPSD retains GT plans for flow supervision; DiT OPSD retains
GT token supervision while distilling rendering under student plans.
These plans determine both DiT conditions and the states queried for
distillation. Joint OPSD adds planning distillation on student prefixes;
all variants update the shared Transformer.
Further reductions in classifier KL and CRW over DiT OPSD support
combining planning and rendering distillation on student prefixes and
visited states, improving content fidelity and spatial execution.

\begin{table}[!htb]
\centering
\small
\setlength{\tabcolsep}{1.8pt}
\renewcommand{\arraystretch}{1.12}
\caption{Editing quality after post-training.
AR OPSD and DiT OPSD retain GT supervision in the other branch.
Bold and underlined values indicate the best and second-best scores per metric.}
\label{tab:opsd-editing}
\label{tab:opsd-ablation}
\begin{tabularx}{\linewidth}{c*{9}{>{\centering\arraybackslash}X}}
\toprule
& \multicolumn{5}{c}{Content quality}
& \multicolumn{4}{c}{Reconstruction and binaural} \\
\cmidrule(lr){2-6}\cmidrule(lr){7-10}
\ambitmetriccell{System}
& \ambitmetriccell{Paired\\CLAP $\uparrow$}
& \ambitmetriccell{FD\\CLAP $\downarrow$}
& \ambitmetriccell{FAD $\downarrow$}
& \ambitmetriccell{FD\\PANNs $\downarrow$}
& \ambitmetriccell{KL $\downarrow$}
& \ambitmetriccell{LSD $\downarrow$}
& \ambitmetriccell{GCC $\downarrow$}
& \ambitmetriccell{CRW $\downarrow$}
& \ambitmetriccell{FSAD $\downarrow$} \\
\midrule
\method{}
& 0.839 & 0.047 & 0.242 & 6.012 & 0.455
& 1.268 & 14.805 & 11.947 & 0.0142 \\
\method{} + SFT
& 0.855 & 0.042 & 0.232 & 5.398 & 0.402
& 1.110 & 11.897 & 7.518 & 0.0119 \\
\method{} + ReFL
& 0.853 & 0.044 & 0.233 & 5.505 & 0.391
& 1.146 & 12.053 & 7.845 & \underline{0.0116} \\
\method{} + GRPO
& 0.851 & 0.043 & 0.228 & 5.553 & 0.417
& 1.149 & 12.553 & 8.923 & 0.0117 \\
\midrule
\method{} + AR OPSD
& 0.852 & 0.042 & 0.237 & 5.472 & 0.413
& 1.109 & 11.903 & 8.193 & 0.0117 \\
\method{} + DiT OPSD
& \underline{0.860} & \underline{0.041} & \underline{0.223}
& \underline{5.314} & \underline{0.385}
& \underline{1.091} & \underline{11.606} & \underline{6.766}
& 0.0117 \\
\rowcolor{blue!10}
\method{} + OPSD
& \textbf{0.867} & \textbf{0.040} & \textbf{0.211}
& \textbf{5.282} & \textbf{0.351}
& \textbf{1.076} & \textbf{11.130} & \textbf{5.854}
& \textbf{0.0115} \\
\bottomrule
\end{tabularx}
\end{table}

\endgroup

\begin{samepage}
\section{Conclusion and Limitations}
\label{sec:conclusion}

We presented \method{}, a unified Transfusion framework for native
FOA generation and editing.
Executable \sceneplan{}s connect source-level content, timing, and
motion to continuous rendering, stating both requested edits and
retained sources; contrastive features ground the change, while full
reference latents supply residual acoustic detail for preservation.
Our FOA-aware VAE combines omnidirectional recovery with
inter-channel supervision, and joint Transfusion OPSD refines
planning and rendering on student-generated contexts.
Experiments across speech, music, and sound show improvements in
content fidelity, spatial accuracy, and unedited-source preservation.
Joint OPSD further improves all nine reported metrics over continued
supervised training, ReFL, GRPO, and both branch supervision variants.
These results support jointly learning scene control and acoustic
execution for spatial audio authoring.

Our current \sceneplan{} configuration supports scenes up to
approximately 15\,s.
Extending \method{} to longer soundscapes and evaluating acoustic
consistency across successive edits remain important directions
for future work.
\end{samepage}

\bibliographystyle{plainnat}
\bibliography{references}

\clearpage
\appendix
\pdfbookmark[0]{Appendix}{appendix.overview}
\begin{center}
  {\LARGE\bfseries Appendix\par}
\end{center}
\raggedbottom
\setlength{\intextsep}{8pt plus 2pt minus 2pt}
\setlength{\textfloatsep}{10pt plus 2pt minus 2pt}
\setlength{\floatsep}{8pt plus 2pt minus 2pt}
\setcounter{topnumber}{2}
\setcounter{bottomnumber}{2}
\setcounter{totalnumber}{4}
\renewcommand{\bottomfraction}{0.8}
\renewcommand{\floatpagefraction}{0.9}

\section{Dataset details}
\label{app:data}

Collecting finely aligned audio data and detailed attribute annotations remains labor-intensive~\citep{zhang2024gtsinger,guo2025stars,pan2025syntheticsingers}.
Table~\ref{tab:sceneplan-schema} lists the \sceneplan{} field ranges
for scene and source attributes
(Section~\ref{sec:formulation}).
Synthetic scenes are rendered at 44.1\,kHz using ACN $[W,Y,Z,X]$
channel order and SN3D normalization.
Coordinates point forward ($+X$), left ($+Y$), and up ($+Z$);
positive azimuth turns left.

\begin{table}[!htb]
\centering\small
\caption{\sceneplan{} field ranges. Times, angles, and distances use seconds, degrees, and metres.}
\label{tab:sceneplan-schema}
\begin{tabularx}{\linewidth}{c>{\raggedright\arraybackslash}X}
\toprule
Field & Representation and constraints\\
\midrule
Duration & At most 648 latent frames at a 1,024-sample hop (approximately 15.05\,s)\\
Room & Dry, moderate, reverberant, or outdoor\\
Sources & One to four plan local source IDs.\\
Content & Source kind, description, and verbatim transcript for speech\\
Activity & Interval $[o_i,f_i)$ with $0\leq o_i<f_i\leq d$\\
Trajectory & Static pose $(\theta,\psi,\rho)$ or linear motion with 2 to 8 keyframes\\
Gain & Recorded in dB; the planning codec decodes a constant $0$\,dB\\
\bottomrule
\end{tabularx}
\end{table}

\paragraph{Real recordings and captioning.}
The generation dataset combines synthetic scenes with a collection
of 80,000 FOA clips from Sphere360~\citep{liu2025omniaudio}
and 3,809 clips from STARSS23~\citep{shimada2023starss23}.
We segment STARSS23 recordings into consecutive 10\,s clips,
discarding shorter tails and cropping available videos to the
corresponding intervals.
Qwen3-Omni~\citep{xu2025qwen3} receives synchronized video and
the omnidirectional $W$ channel as mono audio to produce concise
captions; clips without video use audio alone.
The full FOA signals are retained in their assigned data partitions.
Available STARSS23 event and spatial annotations are converted
into \sceneplan{} fields.
For Sphere360 and unannotated STARSS23 clips, a second Qwen3-Omni
pass expands each caption into source descriptions, activity intervals,
and trajectories under the \sceneplan{} schema.
It retains explicit cues and infers missing times, directions, and
distances as pseudo annotations.
Training plans supervise AR prediction and provide compiled
conditions for DiT rendering.

\paragraph{Synthetic assets and rendering.}
Speech recordings come from LibriTTS~\citep{zen2019libritts},
Hi-Fi TTS~\citep{bakhturina2021hi}, and HiFiTTS-2~\citep{langman2025hifitts};
music and sound assets come from AudioCaps~\citep{kim2019audiocaps},
AudioSet~\citep{gemmeke2017audio}, FSD50K~\citep{fonseca2022FSD50K},
PicoAudio~\citep{xie2024picoaudio}, VGGSound~\citep{chen2020vggsound},
and MusicCaps~\citep{agostinelli2023musiclm}.
Pyroomacoustics~\citep{scheibler2018pyroomacoustics} supplies four-channel
room responses at trajectory keyframes. Source signals are placed at
their onsets, convolved with these responses, and combined by
interpolating adjacent keyframes. Static sources use one response.
Scene gain targets $-23$\,dBFS $W$ channel RMS, subject to a
$-1$\,dBFS true peak ceiling.

For synthetic scene construction, the listener is horizontally centred at
height $\min(1.7,H/2)$\,m.
Initial azimuth, elevation, and distance are sampled from
$[-180,180)^\circ$, $[-22,22]^\circ$, and $[0.9,2.2]$\,m,
with a $0.3$\,m wall margin.
Table~\ref{tab:room-sampling} gives the room distributions used
for scene rendering.

\begin{table}[!htb]
\centering\small
\caption{Room sampling ranges for synthetic scenes.
Outdoor scenes use no reflections.}
\label{tab:room-sampling}
\begin{tabular}{cccc}
\toprule
Class & Dimensions $L\times W\times H$ (m) & RT60 (s) & Image order\\
\midrule
Dry & $[6,8]\times[5,7]\times[3,4]$ & $[0.16,0.28]$ & 5 to 7\\
Moderate & $[8,12]\times[7,10]\times[3.2,4.8]$ & $[0.38,0.62]$ & 6 to 9\\
Reverberant & $[12,18]\times[10,15]\times[4.5,7]$ & $[0.78,1.20]$ & 6 to 9\\
Outdoor & $[10,16]\times[10,16]\times[5,8]$ & --- & 0\\
\bottomrule
\end{tabular}
\end{table}

\paragraph{Editing and partitions.}
Requests express source content and qualitative spatial or temporal
constraints, retaining literal speech transcripts.
Each edit applies \textsc{spatial/time relocate}, \textsc{static$\to$linear},
\textsc{linear$\to$static}, \textsc{add}, or \textsc{remove} to one source.
Dataset annotations retain source correspondence and assets across
an editing pair; serialized plans use the local indexing described
in Appendix~\ref{app:editing}.
Room and scene duration remain fixed, and source and target share
calibration and master gain apart from peak ceiling attenuation.
Synthetic generation partitions are disjoint in source assets;
editing inherits these partitions and draws addition assets from
the same partition.
Sphere360 recordings are used only for generation training and validation.
The 8,000-example generation test set includes 50 FOA clips of
10\,s each from the annotated STARSS23 dev-test split.
These clips are included in the reported generation results.
Splits are assigned by original recording or video, with all
derived clips kept in the same partition.
The original recordings underlying the 50 test clips, including
all other clips derived from those recordings, are excluded from
training and validation.
Real recordings outside these held-out recording groups are used
for generation training and validation, following the same
training-to-validation ratio as synthetic scenes.
Table~\ref{tab:splits} summarizes the overall partitions for both tasks.

\begin{table}[!htb]
\centering\small
\caption{Dataset partitions. Generation training combines synthetic
scenes and real recordings. The 8k generation test examples include
50 FOA clips of 10\,s each from the annotated STARSS23 dev-test split.}
\label{tab:splits}
\begin{tabular}{crrr}
\toprule
Partition & Train & Validation & Test\\
\midrule
Generation examples & 1.6M & 32k & 8k\\
Editing pairs & 1M & 20k & 5k\\
\bottomrule
\end{tabular}
\end{table}

\paragraph{\method{} VAE training data.}
The VAE corpus contains 1M training clips, 20k validation clips,
and 2,000 test clips, combining synthetic spatial audio with FOA
clips from Spatial LibriSpeech~\citep{sarabia2023spatial}.
For music and sound, mono assets are rendered with
Pyroomacoustics~\citep{scheibler2018pyroomacoustics}
under varied room acoustics, source positions, and motion,
then mixed into FOA scenes.
Speech recordings are spatialized into FOA scenes~\citep{zhu2026spatialomni}.
For clips already provided in FOA format, we retain their original channels.

\paragraph{\method{} CLAP training data.}
We train \method{} CLAP on 1M source-target audio pairs from the editing training split.
Each recording is a synthetic FOA scene: music and sound assets are rendered with Pyroomacoustics~\citep{scheibler2018pyroomacoustics} and mixed under a shared room and duration, while speech recordings are spatialized into FOA.
From the paired \sceneplan{}s, we construct a semantic caption of source content and a scene caption that binds each source to its activity, trajectory, and gain.
Relocate and motion edits keep the semantic caption fixed and change the scene caption; additions and removals change both.
We construct up to four negative captions per recording by swapping trajectories or activity intervals between distinguishable sources.

\section{Model details}
\label{app:implementation}
Fine-grained control spans reference conditioning~\citep{zhang2024stylesinger,zhang2024tcsinger,li2026swanvoice}, pretrained generative priors~\citep{zhang2025partassembly}, and sampling guidance~\citep{feng2026rectifying}.
This section specifies the plan interfaces, FOA representation, and
reference conditioning used by the unified model.

\subsection{\sceneplan{} conditioning}
\label{app:compiler}

Frozen Qwen3.5-0.8B features of width 1,024 are projected to
768-dimensional cross-attention tokens for semantic clauses.
The compiler rejects text over 512 tokens to preserve speech.
Each slot concatenates 32 activity and
32 pose embedding channels, giving 256 channels across four sources.
A frame is active whenever its sample
interval overlaps the source activity interval.

At latent frame midpoints clipped to source activity, elevation and
distance interpolate linearly between adjacent keyframes; azimuth
uses the shortest signed increment in $[-\pi,\pi)$.
Endpoint poses extend beyond the keyframe range. Only then do we compute
$(\sin\theta,\cos\theta,\sin\psi,\cos\psi,\log(1+\rho))$
and apply a $5\to64\to32$ MLP with SiLU activations.
Inactive source blocks are zero; dropped conditions use learned unknown
embeddings. Text and gain are excluded from spatial controls.

\subsection{\method{} variational autoencoder}
\label{app:vae}

\paragraph{Architecture and KL reduction.}
The Oobleck based VAE uses base width 128, channel multipliers
$(1,2,4,8,16)$, strides $(2,2,4,8,8)$, and Snake activations.
Its 64 latent channels have a hop of 1,024 samples, with 40 channels
assigned to $z^W$ and 24 to $z^R$.
For posterior scale $\sigma=\operatorname{softplus}(s)+10^{-4}$,
the group penalties in Equation~\ref{eq:vae-w} are computed as follows:
\begin{equation}
 K_g=\frac{1}{BT}\sum_{b,t}\sum_{c\in g}
 \left(\mu_{bct}^2+\sigma_{bct}^2-\log\sigma_{bct}^2-1\right).
 \label{eq:vae-group-kl-app}
\end{equation}
Here $B$, $T$, and $g$ denote batch size, latent sequence length,
and channel group. The reduction sums channels and averages examples
and time steps; coefficients absorb the conventional $1/2$.
The first four ablation configurations use $10^{-4}K_{\mathrm{all}}$
over all 64 channels. The full configuration replaces this uniform
penalty with $\beta_WK_W+\beta_RK_R$, using
$\beta_W=5\times10^{-5}$ and $\beta_R=2\times10^{-4}$.

\paragraph{Reconstruction settings.}
STFT reconstruction uses equally weighted spectral convergence
and log magnitude terms with perceptual K weighting and an overall
coefficient of 1. Hann windows of 32 to 2,048 samples in powers
of two use quarter window hops.
The multi scale STFT discriminator uses adversarial and feature
matching coefficients of 0.1 and 5.0; waveform $L_1$ is disabled.
Phase supervision combines instantaneous frequency, group delay,
and normalized complex distance with unit internal weights
and target weight 0.08.
The full configuration uses cepstral, SI-SDR, and high frequency
overshoot coefficients of 0.15, 0.003, and 0.05.
Auxiliary $W$ recovery reuses the decoder with $z^R$ zeroed,
the configured spectral objective weighted by 0.3, and SI-SDR by 0.003.

\paragraph{Base configuration and cumulative additions.}
Phase (base) includes the multiresolution STFT, within-channel phase,
high frequency overshoot, adversarial, and feature matching terms above,
with uniform KL regularization.
Successive configurations add SCM, cepstral, and SI-SDR losses.
The full configuration retains these terms, replaces uniform KL with
grouped KL, and adds auxiliary $W$ reconstruction.
Thus $\mathcal L_{\mathrm{base}}$ in Equation~\ref{eq:vae-total}
collects the active reconstruction and adversarial terms~\citep{zhu2026csavocoder}, including
cepstral and SI-SDR terms in the full model. Within-channel phase
supervision and SCM's inter-channel phase alignment remain distinct.

\paragraph{Spatial covariance matching.}
For four channel STFT vectors $X(f,\tau)\in\mathbb C^4$,
we compute $C=\langle XX^{\mathsf H}\rangle_5$ over five neighboring
frames, where $\mathsf H$ denotes conjugate transpose.
We use Hann windows, FFT sizes 512 and 2,048, quarter window hops,
and no STFT centering. Temporal averaging excludes padded positions
at sequence boundaries.

Let $e_{brf\tau}=\operatorname{tr}C_{br}(f,\tau)$ denote reference
trace energy for example $b$ and resolution $r$, and let $N_r$ denote the number of time-frequency bins. The shared stabilization floor is
\begin{equation}
 \delta_{br}=\max\!\left(10^{-12},
 \frac{10^{-3}}{N_r}\sum_{f,\tau}e_{brf\tau}\right).
 \label{eq:scm-floor}
\end{equation}
This floor is shared across bins and between prediction and reference
within each example and resolution. Suppressing these indices,
Equation~\ref{eq:scm-main} uses
\begin{equation}
 p_i=\frac{C_{ii}}{\operatorname{tr}C+\delta},\qquad
 \gamma_{ij}=\frac{|C_{ij}|}
 {\sqrt{(C_{ii}+\delta)(C_{jj}+\delta)}},\qquad
 U_{ij}=\frac{C_{ij}}{|C_{ij}|+\delta}.
 \label{eq:scm-statistics}
\end{equation}
Hats use reconstructed covariance; the overline denotes complex
conjugation. Since $|U_{ij}|<1$, SCM's stabilized phase alignment
retains magnitude dependence and need not vanish for identical covariance.

With $\epsilon=10^{-8}$, reference weights are
\begin{equation}
 \begin{aligned}
 w_{brf\tau}&=\frac{e_{brf\tau}}
 {\sum_{f',\tau'}e_{brf'\tau'}+\epsilon},\\
 v_{brij f\tau}&=w_{brf\tau}\,
 \operatorname{clip}(\gamma_{brij f\tau},0,1)
 \left(0.1+\frac{0.9}{1+(f/1500\,\mathrm{Hz})^2}\right).
 \end{aligned}
 \label{eq:scm-weights}
\end{equation}
Here $f$ is physical frequency. These weights are determined by the
reference; the coherence factor is detached. For per bin errors $A$
and pairwise errors $B$, the reductions are
\begin{equation}
 \langle A\rangle_w=
 \frac{\sum_{b,r,f,\tau}w_{brf\tau}A_{brf\tau}}
 {\max(\sum_{b,r,f,\tau}w_{brf\tau},\epsilon)},\qquad
 \langle B\rangle_v=
 \frac{\sum_{b,r,f,\tau}\sum_{i<j}v_{brij f\tau}B_{brij f\tau}}
 {\max(\sum_{b,r,f,\tau}\sum_{i<j}v_{brij f\tau},\epsilon)}.
 \label{eq:scm-reductions}
\end{equation}
Energy and coherence errors sum over four channels and six unordered
pairs, respectively; phase alignment averages all six pairs with
weights favoring coherent, low-frequency observations.
All three internal coefficients are 1; the outer coefficient is
$\lambda_{\mathrm{SCM}}=0.015$.

\subsection{\method{} CLAP reference features}
\label{app:foa-clap}

VAE latents pooled with stride 4, together with mean/RMS features
and positional encodings, enter a Transformer with six layers,
width 384, and six attention heads.
It produces temporal features and normalized semantic and scene
embeddings of dimensions 512 and 256, respectively, aligned with
separate projections of frozen text features.
For head $h\in\{\mathrm{sem},\mathrm{scene}\}$, let $a_i^h$ and
$t_j^h$ denote normalized audio and text embeddings.
Scaled cosine scores $s_{ij}^h=\alpha_h(a_i^h)^\top t_j^h$,
with logit scale $\alpha_h$, give the audio to text loss
\begin{equation}
 \ell_h^{A\to T}
 =-\frac{1}{B}\sum_i\frac{1}{|\mathcal P_h(i)|}
 \sum_{j\in\mathcal P_h(i)}
 \log\frac{\exp(s_{ij}^{h})}
 {\sum_{k\in\mathcal V_h(i)}\exp(s_{ik}^{h})}.
 \label{eq:foa-clap-contrastive}
\end{equation}
Here $B$ is the batch size, $\mathcal P_h(i)$ contains positive
text matches for audio $i$, and $\mathcal V_h(i)$ contains eligible
text candidates, excluding ambiguous partial content overlaps.

Semantic positives match source content and counts.
Scene positives additionally match each source's spatial and
temporal attributes and room properties.
Spatial edits that preserve content are semantic positives
and scene negatives.
Swapped trajectories or activity intervals provide hard negatives
for source attribute binding.

We average the audio to text and text to audio losses for each
head to obtain $\mathcal L_{\mathrm{sem}}$ and
$\mathcal L_{\mathrm{scene}}$.
The complete objective supplements
Equation~\ref{eq:foa-clap-alignment} with binding, editing pair,
and source attribute terms:
\begin{equation}
 \mathcal L_{\mathrm{AMBIT\text{-}CLAP}}
 =\mathcal L_{\mathrm{align}}
 +\lambda_{\mathrm{bind}}\mathcal L_{\mathrm{bind}}
 +\lambda_{\mathrm{pair}}\mathcal L_{\mathrm{pair}}
 +\lambda_{\mathrm{event}}\mathcal L_{\mathrm{event}}.
 \label{eq:foa-clap-total}
\end{equation}
The selected configuration uses $\lambda_{\mathrm{scene}}=1$,
$\lambda_{\mathrm{bind}}=\lambda_{\mathrm{pair}}=5$, and
$\lambda_{\mathrm{event}}=0.1$.
Binding and pair terms rank matched descriptions above attribute
swapped or crossed descriptions; four learned source queries
supervise content and attributes under a shared source assignment.
After pretraining, the encoder is frozen and supplies global
and temporal features $F(z_s)$ to the editing planner.
It is separate from the audio CLAP model used for content
quality evaluation.

\subsection{Generation planning and rendering}
\label{app:generation}
\label{app:planner}

The renderer has 15 Transformer blocks, width 1,024, and 16 attention
heads. Its input concatenates 64 noisy latent channels, 256 spatial
condition channels, and 64 reference latent channels.
Generation sets the reference channels to zero, retaining the same
384 channel input interface used for editing. Speech active frames receive weight
$w_t=3$, compared with $w_t=1$ elsewhere, in the normalized residual loss
for rendering, with frame masks and weights broadcast across channels:
\begin{equation}
 \frac{1}{64\sum_t m_t w_t}\|m\odot\sqrt{w}\odot r\|_2^2
 =\frac{\sum_{c,t}m_t w_t r_{ct}^{2}}
 {64\sum_t m_t w_t}.
 \label{eq:speech-norm}
\end{equation}
For sound-only clips, a loss with weight 0.05 compares latent differences
at lags 1, 2, and 4:
\begin{equation}
 \mathcal L_{\Delta}
 =\frac{1}{3}\sum_{\ell\in\{1,2,4\}}\operatorname{SmoothL1}
 (\Delta_\ell\widehat z_0,\Delta_\ell z_0),\qquad
 \widehat z_0=z_t-tv_\theta.
 \label{eq:temporal-diff}
\end{equation}
Only coordinate pairs valid at both times contribute to
$(\Delta_\ell z)_{:,t}=z_{:,t+\ell}-z_{:,t}$.
Semantic and spatial conditions are independently dropped with
probability 0.15 for each training example.

\paragraph{Autoregressive plan completion.}
The planner converts raw requests into executable \sceneplan{}s for
speech, music, environmental sounds, and their combinations.
Inventory, span selection, and qualitative heads determine source
content, motion, and activity; completed fields condition subsequent
AR tokens. Table~\ref{tab:planner-field-ownership} specifies which
fields use head predictions, numerical completion, or AR logits.

The codec has a vocabulary of 4,096 tokens, including 1,536 text pieces,
649 frame values, $1^\circ$ angular bins, and 256 logarithmic distance
bins over 0.1 to 50\,m. Onsets are rounded down and offsets are rounded up to the latent-frame grid. Grammar constrained greedy decoding serializes scene
and source fields.
Planning LoRA has rank 8 and $\alpha=8$ and is active only in AR mode.

\begin{table}[!htbp]
\centering
\footnotesize
\setlength{\tabcolsep}{3pt}
\renewcommand{\arraystretch}{1.12}
\caption{Source and scene fields completed by the AR planner.}
\label{tab:planner-field-ownership}
\begin{tabularx}{\linewidth}{
 >{\centering\arraybackslash}p{0.19\linewidth}
 >{\raggedright\arraybackslash}p{0.21\linewidth}
 >{\raggedright\arraybackslash}X}
\toprule
Plan fields & AR mechanism & Prediction and completion \\
\midrule
Room and duration
& AR logits
& Four room classes and 1 to 648 duration frames. \\
Source inventory
& Inventory head
& Predict source count and sound, music, or speech categories. \\
Description and transcript
& Learned span selectors
& Select literal request spans and encode them with the text codec. \\
Motion and azimuth
& Qualitative head
& Static or linear; eight compass categories for start and end directions. \\
Onset and offset
& Qualitative head
& Five relative time phases; select a feasible phase pair and complete integer frame values. \\
Radial motion
& Qualitative head
& Free, approaching, or receding; complete endpoint distances. \\
Structural tokens and gain
& Deterministic rules
& Grammar supplies required delimiters; decoded gain is fixed at 0\,dB. \\
\bottomrule
\end{tabularx}
\end{table}

\paragraph{Spatial and temporal completion.}
Generation instantiates qualitative controls deterministically using
seed 42 and request, source, and field keys. These defaults differ
from the data sampling ranges in Appendix~\ref{app:data}.
Eight compass centres start at front ($0^\circ$) and advance leftward
in $45^\circ$ steps. Azimuths lie within $8^\circ$ of the selected
centre before codec quantization.
Initial distance lies in $[2.2,3.25)$\,m.
Approaching and receding linear sources decrease or increase it by
$[0.75,1)$\,m; free radial motion keeps distance fixed.
Static sources use one position selected by summed endpoint scores.
Coincident linear endpoints receive a $6^\circ$ azimuth offset.

Beginning, early, middle, late, and ending use duration fractions
$[0,0.2]$, $[0.1,0.4]$, $[0.3,0.7]$, $[0.6,0.9]$, and
$[0.8,1]$, respectively. The highest scoring feasible phase pair
determines integer onset and offset frames, with onset before offset
and a minimum gap of ten frames when feasible.
Speech uses values within two frames of the feasible outer boundaries.
These values fix numerical tokens in the AR stream; editing prediction
is detailed in Appendix~\ref{app:editing}.

\subsection{Editing planner and renderer}
\label{app:editing}

Editing fills the 64 reference channels with source audio latents,
using the same input projection and Transformer blocks as generation.
The projection columns associated with source latents are initialized
to zero and learned during training.
The AR prefix combines native latents with global and temporal
\method{} CLAP features:
\begin{equation}
 h_{s,\ell}=h^{\mathrm{base}}_{s,\ell}
 +P_g[g_{\mathrm{sem}};g_{\mathrm{scene}}]
 +[\operatorname{Up}_4(P_{\mathrm{temp}}H)]_\ell.
 \label{eq:bridge}
\end{equation}
These projections start at zero; $\operatorname{Up}_4$ repeats features
over four latent frames. A two-layer decoder with source slots and a
\textsc{new} slot supplies AR memory through gated residual attention.

\paragraph{Autoregressive edit completion.}
Source audio and raw edit instructions condition a complete target
\sceneplan{}, describing retained and modified sources with their
content, spatial attributes, and onset and offset times.
Predicted source attributes and edit decisions supply memory for AR
decoding, connecting a local request to the full scene rendered by DiT.
Duration follows the source audio, and decoded gain remains 0\,dB.

\paragraph{Canonical target order.}
Targets are stably sorted by codec projected onset, source kind, and
compact JSON of source content and attributes with sorted keys.
Exact ties retain input order and have identical projected content.
Reordering preserves acoustic fields and paired source correspondences,
aligning AR targets, rendering conditions, and edit annotations without
changing source identity. Inference uses the learned order.

\paragraph{Shared source supervision.}
Equation~\ref{eq:edit-joint} uses $\lambda=0.1$.
Paired annotations link source attributes before and after editing.
We minimize the field cost over injective assignments from
occupied source records to predicted slots, including the kind cost of
unused slots. Cost ties use the first assignment in
lexicographic slot order.
One assignment supervises recognition, edit binding, preservation,
modification, and target presence, without rematching after editing.

Removals keep their assigned slot with \textsc{remove} and an empty
target; additions use the \textsc{new} slot and enter canonical order.
Field masks distinguish preserved attributes from requested changes.

\section{Training and inference settings}
\label{app:training}

\method{} supports FOA generation and editing within a single
Transfusion model, sharing Transformer blocks across autoregressive
planning and rectified flow rendering.
The complete student contains 332.6M trainable parameters,
including the shared backbone, native task interfaces, and
structured heads.
Generation uses language instructions, while editing additionally
conditions on source audio.
Table~\ref{tab:training-settings} summarizes the learning rates
for VAE pretraining and joint Transfusion optimization.
We train \method{} on eight NVIDIA L40S GPUs.

\paragraph{\method{} VAE pretraining.}
\method{} VAE is trained on approximately 1M FOA clips, including
synthetic speech and sound scenes and Spatial
LibriSpeech~\citep{sarabia2023spatial}, using 4\,s crops.
All cumulative VAE ablations in Table~\ref{tab:vae-all}
are trained from scratch for matched optimization steps
on the same data split.

\paragraph{Joint Transfusion optimization.}
Generation and editing use the same Transfusion model,
with shared Transformer blocks for AR planning and DiT rendering.
Generation training uses construction plans for synthetic scenes
and converted or caption derived \sceneplan{} annotations for
real recordings, as described in Appendix~\ref{app:data}.
During generation planner warm-up, the pretrained renderer
remains frozen while planning adapters, heads, and
LoRA~\citep{hu2021lora} parameters are trained;
LoRA is active only during planning.
Subsequent joint optimization on editing pairs uses the paired
objective in Section~\ref{sec:transfusion}, updating the shared
Transformer, native adapters, and structured heads.
Paired rendering uses compiled ground truth target plans,
whereas inference uses compiled student predictions.
For editing, source latents condition both modes, while
\method{} CLAP features additionally condition the planner.
The renderer's source latent input projection columns and
CLAP feature projections are initialized to zero.
The pretrained text backbones, \method{} CLAP, and \method{} VAE
remain frozen and are excluded from the trainable parameter count.

\begin{table}[!htb]
\centering
\small
\setlength{\tabcolsep}{4pt}
\renewcommand{\arraystretch}{1.1}
\caption{Learning rates for VAE pretraining and the joint
optimization stage. All updates use AdamW.}
\label{tab:training-settings}
\begin{tabularx}{\linewidth}{
    >{\centering\arraybackslash}p{0.23\linewidth}
    >{\raggedright\arraybackslash}p{0.25\linewidth}
    >{\raggedright\arraybackslash}X}
\toprule
Component & Optimization objective & Learning rates\\
\midrule
\method{} VAE
& FOA reconstruction
& VAE $1.5\times10^{-4}$;
  discriminator $3\times10^{-4}$\\
\midrule
Transfusion
& Joint AR and DiT training
& Adapters and heads $10^{-4}$;
  shared blocks $2\times10^{-6}$\\
\bottomrule
\end{tabularx}
\end{table}

\paragraph{Flow time sampling and inference.}
For generation training, flow times follow
$t=1-(u-0.075)/0.925$, where
$u=\operatorname{sigmoid}(\xi)$ and
$\xi\sim\mathcal N(0,1)$, conditioned on $u\geq0.075$.
Paired editing training samples $t$ uniformly from $[0,1]$~\citep{guo2025techsinger}.

The unified model supports both tasks through AR planning,
deterministic plan compilation, DiT rendering with Euler
integration~\citep{zhang2025versband}, and VAE decoding.
Generation omits source audio conditioning.
Editing uses source latents in both planning and rendering,
with additional \method{} CLAP features for planning
(Section~\ref{sec:transfusion}).
Generation uses 100 Euler steps, guidance scale 3.0,
and guidance rescaling 0.4; editing uses 20 Euler steps
and guidance scale 1.0.

Within each task, planner comparisons hold the VAE and renderer
weights, sampler settings, and noise seeds fixed.
All compared planners produce plans for the same deterministic
compiler and DiT rendering interface.
OPSD further refines the shared student through joint planning
and rendering supervision; its optimization settings and
post-training controls appear in Appendix~\ref{app:opsd}.

\section{Evaluation details and additional results}
\label{app:exp-setup}

\subsection{Evaluation protocol and baseline interfaces}
\label{app:generation-quality-cohorts}
\label{app:editing-quality-cohorts}

Generation and editing results use the 8k generation and 5k editing
benchmarks, respectively. The \method{} VAE is evaluated on 2,000 FOA clips. Music, Sound, and Speech categories can
overlap; pooled spatial results count each scene once.
Audio comparisons use executable plans and metric valid observations
from the relevant benchmark; missing measurements are excluded rather
than assigned zero error.
External generators receive semantic descriptions, while \method{}
receives qualitative requests. Editing baselines use source audio and
an instruction or target description according to their native
interfaces. Qwen3-Omni is a planner for our DiT, distinct from the Audio-Omni editor: its source observation is the $W$ waveform, whereas
\method{} uses native FOA features.

\subsection{External audio spatialization}
\label{app:spatial-cascades}

For BEWO + AS and TangoFlux + AS in Table~\ref{tab:spatial-generation},
Qwen3.5-4B supplies the same executable plans to both systems.
Each audio backend synthesizes source signals from the plan's semantic
fields; these signals are downmixed to mono and placed within their
predicted activity intervals. Pyroomacoustics renders the predicted
positions and room classes in ACN/SN3D FOA, using fixed inference
geometry instead of ground truth room dimensions.
Moving sources crossfade room responses at 250ms intervals.
Qwen3.5-4B + DiT instead passes its plan through our compiler and
neural renderer. These comparisons assess complete planning and
rendering systems with a common target waveform.

\subsection{Metrics}
\label{app:metrics}

The generation, editing, and VAE test sets contain 8,000 requests,
5,000 requests, and 2,000 clips, respectively.
All systems use the same evaluation examples for each task and metric.
Directional errors use a common set of frames with valid direction
estimates in the reference and every compared output.
Object metrics additionally use isolated source intervals.

\paragraph{Content and speech.}
Paired CLAP measures similarity between output and reference audio.
FAD and feature based Fr\'echet distances compare embedding
distributions; PANNs/PaSST KL compares predicted acoustic classes.
Distributional scores use the evaluated features directly.
Generation uses native $W$ or arithmetic mean mono, peak normalized
to $-1$\,dBFS before resampling; editing uses mono downmixes of
binauralized FOA without gain normalization.
CLAP, PANNs, and VGGish operate at 48, 32, and 16\,kHz.
For clips longer than 10\,s, CLAP window embeddings are normalized,
averaged, and renormalized. Distil-Whisper large-v3 supplies
transcriptions; generation WER/CER aggregate edit counts over
reference lengths~\citep{zhang2025conan}, while UTMOS averages individual clip scores.

\paragraph{Reconstruction and binaural quality.}
FOA is decoded with fixed MIT KEMAR filters using eight virtual
speakers at azimuths $45,135,225,315^\circ$ and elevations
$\pm40^\circ$; native stereo is unchanged.
LSD is the framewise RMS difference in log$_{10}$ power at 24\,kHz,
with a 1,024 sample Hann window and 240 sample hop.
GCC is $1,000$ times the squared difference of clip mean interaural
delays in milliseconds, estimated by GCC-PHAT at 16\,kHz with a
$\pm1$\,ms search. CRW compares StereoCRW delay predictions;
FSAD compares the distributions of its spatial audio features.

\paragraph{Scene direction and angular trajectories.}
\label{app:scene-spatial}
The scene columns in Tables~\ref{tab:spatial-generation} and
\ref{tab:spatial-editing} compare output and GT waveform directions.
Active intensity directions use 1,024 sample frames at 44.1\,kHz.
Valid frames require coherence at least 0.1, a nonzero direction,
and FOA energy at least $\max(10^{-12},10^{-4}e_{\max})$ within each
waveform. We intersect valid frames across GT and compared outputs;
editing additionally restricts frames to target plan activity.
For spherical angular difference $d_t$, wrapped azimuth difference
$\Delta\theta_t\in[-\pi,\pi)$, and normalized GT energy weights $a_t$,
\begin{equation}
 \mathrm{DOA}=\sum_t a_td_t,\qquad
 \mathrm{AzMAE}=\sum_t a_t|\Delta\theta_t|,\qquad
 \mathrm{AngRMSE}=\sqrt{\sum_t a_td_t^2}.
 \label{eq:editing-spatial-metrics-app}
\end{equation}
Here $d_t$ is the arccosine of the clipped inner product between
output and GT unit directions. Azimuth excludes undefined horizontal
directions and renormalizes the weights.
Overlapping sources contribute mixture directions.
Errors are in radians and scenes are averaged equally.
Angular trajectory RMSE measures directional deviations over time
without evaluating source distance.

\paragraph{Object direction and angular trajectories.}
\label{app:object-spatial}
The object columns use scenes with one to four GT sources.
For each object, eligible frames contain exactly one active GT source;
its interval mask is eroded by two frames at each boundary.
Intersecting this mask with common valid direction frames gives
$\mathcal J_{so}$ for object $o$ in scene $s$.
With GT energy weights $a_{so,t}$ normalized over this interval,
\begin{equation}
 \mathrm{SphMAE}_{so}
 =\sum_{t\in\mathcal J_{so}}a_{so,t}d_t,\quad
 \mathrm{AzMAE}_{so}
 =\sum_{t\in\mathcal J_{so}}a_{so,t}|\Delta\theta_t|,\quad
 \mathrm{AngRMSE}_{so}
 =\sqrt{\sum_{t\in\mathcal J_{so}}a_{so,t}d_t^2}.
 \label{eq:object-spatial-metrics}
\end{equation}
Azimuth excludes frames whose horizontal unit direction norm is at
most $10^{-6}$ in either waveform and renormalizes the remaining
weights. Object metrics compare actual output and GT FOA directions;
they do not compare output directions with the predicted plan.
For any object metric $m_{so}$, the reported average is
\begin{equation}
 \overline m=\frac{1}{|\mathcal V|}\sum_{s\in\mathcal V}
 \frac{1}{|\mathcal O_s|}\sum_{o\in\mathcal O_s}m_{so},
 \label{eq:object-spatial-aggregation}
\end{equation}
where $\mathcal O_s$ and $\mathcal V$ denote valid objects and scenes
for each metric. Frames use GT energy weights. We weight valid objects equally
within each scene and then average equally over all eligible scenes
with one to four sources. Object RMSE is computed per object before
this averaging.
We apply no source separation, temporal alignment, or angular
reassignment. The exclusive activity mask can retain reverberant
tails despite its boundary guard.

\paragraph{Spatial feature similarity.}
We use PSELDNets activity coupled direction features at 24\,kHz.
For vectors $a_{tc}$ and $b_{tc}$ from output and reference,
\begin{equation}
 \mathrm{wCS}=
 \frac{\sum_{t,c}a_{tc}^{\mathsf T}b_{tc}}
 {\sum_{t,c}\|a_{tc}\|_2\|b_{tc}\|_2}.
 \label{eq:wcs-app}
\end{equation}
This magnitude weighted cosine uses all 170 classes without
thresholding or class selection. Padding frames are excluded;
a zero denominator is treated as an undefined measurement.

\paragraph{Unedited source preservation.}
The preservation block of Table~\ref{tab:spatial-editing} compares
output with source audio where the same unchanged source is the only
annotated active source in both source and target plans.
The mask excludes the union of edited source activity in both plans,
then removes two frames from each retained interval boundary.
For time relocation, both original and shifted intervals are excluded;
for removal, the original interval remains excluded.
Direction MAE uses common valid direction frames and source energy
weighting. $W$ NMSE divides squared waveform error
by source energy over eligible samples; absolute level change is
$|10\log_{10}(P_{\mathrm{out}}/P_{\mathrm{src}})|$,
where $P$ denotes mean $W$ power on those samples.
These masks establish exclusivity from activity annotations,
without acoustic source separation; reverberant tails may remain
despite the boundary guard.

\paragraph{VAE comparison.}
\label{app:vae-protocol}
\label{app:vae-ablation}
Table~\ref{tab:vae-all} reports codec reconstruction and cumulative
objective ablations. The ablations follow
Appendices~\ref{app:vae} and~\ref{app:training}, with a 24\,kHz
evaluation frontend.
$W$ SI-SDR and paired CLAP assess omnidirectional reconstruction;
direction and intensity coherence errors assess spatial reconstruction.
Spatial scoring uses FOA signals resampled to 24\,kHz after a
single delay estimated from $W$ by GCC-PHAT is applied to all
channels, retaining only their common temporal overlap.

\paragraph{Intensity coherence error.}
The development of this metric is inspired by FOA tokenizer~\citep{sudarsanam2025foa} and DirAC~\citep{pulkki2006directional}. For each complete 40\,ms frame with a 20\,ms hop, let
$\mathbf d=(Y,Z,X)^{\mathsf T}$ and let $\langle\cdot\rangle_k$
denote the uniform sample mean in frame $k$, without subtracting
the channel means.
Define $e_k^W=\langle W^2\rangle_k$,
$e_k^D=\langle\|\mathbf d\|_2^2\rangle_k$, and
$\mathbf i_k=\langle W\mathbf d\rangle_k$.
Coherence and its error are
\begin{equation}
 r_k=\operatorname{clip}_{[0,1]}\!\left(
 \frac{\|\mathbf i_k\|_2}
 {\sqrt{\max(e_k^We_k^D,10^{-16})}}\right),\qquad
 E_{\mathrm{IC}}(x,\widehat x)
 =\frac{\sum_k a_k E_k\,|\widehat r_k-r_k|}
 {\max\!\left(\sum_k a_k E_k,10^{-12}\right)}.
\label{eq:intensity-coherence-error}
\end{equation}
Here $E_k=e_k^W+e_k^D$ is the total reference channel energy.
A frame is active when $E_k>\max(10^{-3}\max_l E_l,10^{-10})$;
$a_k$ is its binary activity mask.
Hatted statistics are computed independently from the reconstruction;
the mask and aggregation weights use only the reference.
We average $E_{\mathrm{IC}}$ equally across clips.
This time domain statistic measures the magnitude of the normalized
intensity vector, whereas SCM uses pairwise STFT covariance statistics.
Direction error uses the angle between reference and reconstructed
intensity vectors in degrees, weighted by $\|\mathbf i_k\|_2$
over active reference frames with $r_k\geq0.1$.
The additional threshold applies only to direction error;
intensity coherence error uses all active reference frames,
including frames of low coherence.

\subsection{Subjective listening study}
\label{app:user-study}

\paragraph{Participants and consent.}
We recruited 20 student participants from our university.
All participants provided informed consent before the study,
including consent to record their audio preferences and ratings
for research purposes.

\paragraph{Cases and assignment.}
We prepared separate pools of 1,000 matched FOA cases for generation
and editing, and randomly partitioned each pool into 50 non-overlapping
sets of 20 cases. Each participant selected one available set per task
without previewing its contents; no set was reused. Each case contained
three system outputs for the same request. The generation panel compared
\method{}, BEWO + AS, and TangoFlux + AS; the editing panel compared
\method{}, SwanWeave, and Qwen3-Omni + DiT. The study covered 400 distinct
cases per task and 2,400 system-output presentations in total, with one
participant rating all three outputs for each case.

\begin{table}[!htb]
\centering
\footnotesize
\setlength{\tabcolsep}{3pt}
\caption{Subjective evaluation of FOA generation and editing.
All dimensions use a five-point scale; higher is better.
Entries report mean ratings over valid responses.
The highest mean in each column is highlighted in bold.}
\label{tab:user-study}

\textbf{(a) Generation.}\par\vspace{3pt}
\begin{tabularx}{\linewidth}{c*{4}{>{\centering\arraybackslash}X}}
\toprule
System
& \shortstack{Quality\\MOS}
& \shortstack{Instruction\\adherence}
& \shortstack{Spatial\\adherence}
& \shortstack{Spatial\\naturalness} \\
\midrule
BEWO + AS
& 2.93 & 2.96 & 2.83 & 2.89 \\
TangoFlux + AS
& 2.77 & 2.84 & 2.63 & 2.76 \\
\rowcolor{blue!10}
\method{}
& \textbf{3.11} & \textbf{3.25} & \textbf{3.37} & \textbf{3.03} \\
\bottomrule
\end{tabularx}

\par\vspace{8pt}
\textbf{(b) Editing.}\par\vspace{3pt}
\begin{tabularx}{\linewidth}{c*{3}{>{\centering\arraybackslash}X}}
\toprule
System
& \shortstack{Quality\\MOS}
& \shortstack{Edit\\success}
& \shortstack{Unedited content\\preservation} \\
\midrule
SwanWeave
& 1.89 & 1.96 & 1.93 \\
Qwen3-Omni + DiT
& 3.51 & 3.26 & 3.44 \\
\rowcolor{blue!10}
\method{}
& \textbf{3.64} & \textbf{3.57} & \textbf{3.71} \\
\bottomrule
\end{tabularx}
\end{table}

\paragraph{Listening procedure.}
Participants used headphones after a brief familiarization session.
System identities were concealed, and presentation order was randomized
within each case. All FOA outputs used the same fixed MIT KEMAR binaural
decoder described in Appendix~\ref{app:metrics}, with a shared playback
gain within each case. Generation trials presented the original request;
editing trials additionally provided the original audio and editing
instruction. Participants could replay audio and take breaks.

\paragraph{Ratings and reporting.}
Participants rated generation on overall audio quality~\citep{pan2025mesa}, instruction
adherence, spatial adherence, and spatial naturalness. They rated editing
on overall audio quality, edit success, and preservation of unedited
content. Each dimension used a five-point scale from 1 (very poor) to
5 (excellent). Table~\ref{tab:user-study} reports the mean of valid
ratings for each system and dimension. The same participant rated all
three outputs for each case under the same request and playback setup,
enabling within-case comparison.

\paragraph{Results.}
\method{} outperforms both baselines in mean listener ratings across
all seven measures. For generation, it achieves scores of 3.11 in
overall quality, 3.25 in instruction adherence, 3.37 in spatial
adherence, and 3.03 in spatial naturalness. Its largest advantage is
in spatial adherence, where it exceeds the strongest baseline by
0.54 points, followed by a 0.29-point gain in instruction adherence.
For editing, \method{} achieves 3.64 in overall quality, 3.57 in
edit success, and 3.71 in preservation of unedited content, exceeding
Qwen3-Omni + DiT by 0.13, 0.31, and 0.27 points, respectively.
These results highlight \method{}'s strengths in following spatial
instructions and carrying out requested edits while preserving
the surrounding audio content.

\subsection{Planner validity}
\label{app:ar-evaluation}
\label{sec:ar-evaluation}

Table~\ref{tab:ar-plan-evaluation} separates structural planning accuracy
from rendered audio quality. Executability checks schema and compiler
compatibility; source count accuracy checks the requested inventory.
Qwen uses its first response without repair or retries, whereas
\method{} uses grammar-constrained decoding.
All requests in each evaluation panel remain in the denominator,
including failed outputs. Neither measure establishes full instruction
satisfaction or correct source attribute binding.

\begin{table}[!htb]
\centering\small
\setlength{\tabcolsep}{10pt}
\caption{Planner executability and source count accuracy (\%).}
\label{tab:ar-plan-evaluation}
\begin{tabular}{ccc}
\toprule
Planner & Executable $\uparrow$ & Source count accuracy $\uparrow$\\
\midrule
\multicolumn{3}{c}{\textit{Generation}}\\
\midrule
\method{} AR & 100.00 & 100.00\\
Qwen3.5-0.8B & 22.88 & 22.69\\
Qwen3.5-2B & 20.19 & 20.38\\
Qwen3.5-4B & 98.08 & 98.06\\
Qwen3.5-9B & 87.81 & 87.83\\
Qwen3.5-27B & 100.00 & 100.00\\
\midrule
\multicolumn{3}{c}{\textit{Editing}}\\
\midrule
\method{} AR & 100.00 & 86.10\\
Qwen3-Omni & 97.00 & 46.50\\
\bottomrule
\end{tabular}
\end{table}

\begingroup
\setlength{\abovedisplayskip}{4pt plus 1pt minus 1pt}
\setlength{\belowdisplayskip}{4pt plus 1pt minus 1pt}
\setlength{\abovedisplayshortskip}{2pt plus 1pt}
\setlength{\belowdisplayshortskip}{3pt plus 1pt minus 1pt}
\makeatletter
\renewcommand\subsection{\@startsection{subsection}{2}{\z@}%
  {-8pt plus -2pt minus -1pt}{4pt plus 1pt}%
  {\normalfont\normalsize\bfseries\raggedright}}
\renewcommand\paragraph{\@startsection{paragraph}{4}{\z@}%
  {5pt plus 1pt minus 1pt}{-1em}{\normalfont\normalsize\bfseries}}
\makeatother
\section{Transfusion OPSD}
\label{app:joint-refinement}
\label{app:opsd}

We apply and evaluate OPSD as a post-training stage on editing pairs.
Joint OPSD distills both modes on student-generated contexts.
AR OPSD and DiT OPSD replace distillation in the other mode with paired
GT supervision. The following sections specify teacher preparation,
target construction, shared updates, and comparison settings.

\subsection{Teacher construction and reference tokens}
\label{app:opsd-teachers}

Both teachers start from the same supervised model and are frozen after
preparation. The student retains its initial supervised weights;
a separate frozen copy supplies ordinary-condition reference distributions.

\paragraph{Planning teacher.}
The AR teacher inserts complete target plan memory
$M(\mathcal S^\star)$ between source and student tokens.
Source queries read source tokens only; memory queries read source
and memory tokens; student queries additionally read their causal
prefix. Source latents, source \method{} CLAP features, instructions,
and ordinary rotary positions are preserved.
Preparation combines ordinary and target memory token supervision
with paired rendering:
\begin{equation}
 \mathcal L_{\mathrm{prep}}^{\mathrm{AR}}
 =a\mathcal L_{\mathrm{ord}}+b\mathcal L_{\mathrm{mem}}
 +c\mathcal L_{\mathrm{RF}}^{\star}.
 \label{eq:opsd-ar-preparation}
\end{equation}
The token terms use GT prefixes and equally weight mean token CE and
the mean of present field means: source count, kind, text, activity,
motion, and geometry. Rendering uses GT target plans.
Distillation then queries the teacher on the student's generated prefix.

\paragraph{Rendering teacher.}
Clean target latents supply temporally aligned reference tokens:
\begin{equation}
 H_y[\tau]=W_{\mathrm{ref}}z_y[\tau]+r_{\mathrm{ref}},
 \qquad z_y=E_V(x_y),
 \label{eq:opsd-audio-reference}
\end{equation}
where $W_{\mathrm{ref}}$ projects 64 channels to width 1,024 and
$r_{\mathrm{ref}}$ identifies the reference role.
All valid frames are inserted after native prepended tokens, with
rotary times aligned to state frames. Reference queries read reference
tokens only; ordinary queries can additionally read them.
Reference outputs are removed before native post-processing.
The preparation objective uses the mask and normalization of
Equation~\ref{eq:rf}:
\begin{equation}
 \mathcal L_{\mathrm{prep}}^{\mathrm{DiT}}
 =\mathbb E\!\left[
 \frac{1}{\|m\|_1}\big\|m\odot\big(v_{\phi_{\mathrm{DiT}}}(\zeta_t,t;\mathcal S^\star,z_s,H_y)
 -(\epsilon-z_y)\big)\big\|_2^2\right],\quad
 \zeta_t=(1-t)z_y+t\epsilon.
 \label{eq:opsd-teacher-preparation}
\end{equation}
Preparation updates native teacher parameters, the reference projection,
and the role embedding. During refinement, only the teacher receives
these target tokens; student rollouts use ordinary source conditioning.

\paragraph{Preparation configuration.}
Table~\ref{tab:opsd-teacher-settings} specifies preparation budgets
of 1,000 AR updates and 3,000 DiT updates, excluding the shared
supervised initialization. Both use AdamW with a global batch of
128 requests from the editing training partition.
The pretrained text backbones, \method{} CLAP, and VAE remain frozen
during teacher preparation and subsequent student refinement.

\begin{table}[!htbp]
\centering
\footnotesize
\setlength{\tabcolsep}{3pt}
\renewcommand{\arraystretch}{1.10}
\caption{Teacher preparation budgets and settings. Batches are global.}
\label{tab:opsd-teacher-settings}
\begin{tabularx}{\linewidth}{c c c >{\centering\arraybackslash}X c}
\toprule
Teacher & Update budget & Batch & Learning rates & Loss weights \\
\midrule
AR & 1,000 & 128
& $5\times10^{-6}$ / $3.75\times10^{-7}$
& $(a,b,c)=(0.1,0.3,0.5)$ \\
DiT & 3,000 & 128
& $3\times10^{-6}$ / $10^{-4}$
& RF: $1$ \\
\bottomrule
\end{tabularx}
\end{table}

AR learning rates apply to adapters and structured heads, then shared
blocks and rendering modules; DiT rates apply to native parameters,
then the reference projection and role embedding.
DiT uses AdamW betas $(0.9,0.999)$, weight decay $10^{-3}$,
and gradient clipping at norm 1. AR uses its terminal checkpoint.
DiT evaluates checkpoints every 250 updates on 500 fixed validation
requests and selects the checkpoint with the lowest paired RF error
under target audio conditioning.

\subsection{Planning targets and reference constraints}
\label{app:opsd-planning}

Each request produces one grammar constrained greedy plan from source
audio and instructions. Teacher and student distributions use temperature
1 on the same legal support, excluding singleton decisions.
For present decision families $\mathcal F$, indices $\mathcal I_f$,
and detached eligibility $b_i$, the weights are
\begin{equation}
 \eta_i=\frac{b_i}{|\mathcal F|\,|\mathcal I_{f(i)}|}.
 \label{eq:opsd-family-weights}
\end{equation}
Families distinguish count, content, activity, motion, geometry, room,
and other decisions. Room uses ordinary reference supervision.
For unambiguously aligned facts $a_i^\star$, exclude illegal sites or those with
$\max_a q_i(a)-q_i(a_i^\star)>\delta_{\mathrm{fact}}$, with
$\delta_{\mathrm{fact}}=0.05$.
Labels select sites without changing teacher probabilities.
The frozen supervised copy supplies $q_i^{\mathrm{ref}}$ on the same
student prefixes under ordinary conditions for room, event kind,
text, and activity:
\begin{equation}
 \ell_{\mathrm{ref}}
 =\frac{1}{|\mathcal F_{\mathrm{ref}}|}
 \sum_{f\in\mathcal F_{\mathrm{ref}}}\frac{1}{|\mathcal I_f|}
 \sum_{i\in\mathcal I_f}b_i^{\mathrm{ref}}
 D_{\mathrm{KL}}(q_i^{\mathrm{ref}}\|p_{\theta,i}).
 \label{eq:opsd-reference-retention}
\end{equation}
Here $\mathcal F_{\mathrm{ref}}$ contains the present reference families.
Reference eligibility uses the same checks; rejected sites retain the
original denominators. Source indices alone do not establish identity.


\subsection{Student rollouts and rendering targets}
\label{app:opsd-rendering}

Algorithm~\ref{alg:joint-opsd} summarizes one minibatch update.
Student plans determine both the rendering conditions and the visited
states at which the rendering teacher is queried.
For each pair, $c=(z_s,F(z_s),e)$ denotes ordinary editing conditions,
and $H_y=P_{\phi_{\mathrm{DiT}}}(z_y)$ contains target-audio reference tokens
(Equation~\ref{eq:opsd-audio-reference}).
DiT OPSD uses the same student-plan-conditioned rendering procedure,
with its paired GT-prefix planning supervision specified in
Appendix~\ref{app:opsd-variants}.

\begin{algorithm}[!htbp]
\caption{Joint Transfusion OPSD}
\label{alg:joint-opsd}
\small
\begin{algorithmic}[1]
\Require Minibatch $\mathcal B$; student $\theta$; frozen teachers
$\phi_{\mathrm{AR}},\phi_{\mathrm{DiT}}$ and reference model $\theta_{\mathrm{ref}}$
\For{each pair $b=(x_s,e,\mathcal S^\star,x_y)\in\mathcal B$}
  \State $(z_s,z_y,m)\gets\Call{EncodePair}{x_s,x_y}$
  \State $(\widehat{\mathcal S},\{h_i\})
  \gets\operatorname{sg}[\Call{GreedyPlan}{\theta,c}]$
  \State $\{q_i,q_i^{\mathrm{ref}},\eta_i,\eta_i^{\mathrm{ref}}\}
  \gets\operatorname{sg}[\Call{PlanningTargets}{c,\{h_i\},\mathcal S^\star}]$
  \State $\{(Z_j,t_j)\}_{j\in\mathcal Q}
  \gets\operatorname{sg}[\Call{RolloutAndQuery}{\theta,\widehat{\mathcal S},z_s,m}]$
  \Comment{Student plan drives DiT}
  \State $v_j^T\gets\operatorname{sg}
  [v_{\phi_{\mathrm{DiT}}}(Z_j,t_j;\widehat{\mathcal S},z_s,H_y)]$,
  $j\in\mathcal Q$
  \State $p_{\theta,i}\gets p_\theta(\cdot\mid h_i,c)$,
  $i\in\mathcal I$ \Comment{Recompute with gradients}
  \State $v_j^S\gets v_\theta(Z_j,t_j;\widehat{\mathcal S},z_s)$,
  $j\in\mathcal Q$
  \State $\mathcal L_b\gets
  \lambda_{\mathrm{AR}}\ell_{\mathrm{AR}}
  +\lambda_{\mathrm{DiT}}\ell_{\mathrm{DiT}}
  +\lambda_{\mathrm{ref}}\ell_{\mathrm{ref}}$
\EndFor
\State $\mathcal L_{\mathcal B}\gets
|\mathcal B|^{-1}\sum_{b\in\mathcal B}\mathcal L_b$
\State $\theta\gets\Call{AdamWStep}{\theta,\nabla_\theta\mathcal L_{\mathcal B}}$
\Comment{One shared update}
\end{algorithmic}
\end{algorithm}

\textsc{EncodePair} applies the frozen VAE to the source and target
audio, returning their latents $z_s,z_y$ and a binary mask $m$ marking
valid latent coordinates.
\textsc{GreedyPlan} performs grammar-constrained greedy decoding under
$c$: it selects the highest-scoring legal token at each step and returns
$\widehat{\mathcal S}$ and its prefixes $\{h_i\}_{i\in\mathcal I}$ at
non-forced decision positions.
\textsc{PlanningTargets} returns the teacher and reference distributions
$q_i,q_i^{\mathrm{ref}}$ and their weights $\eta_i,\eta_i^{\mathrm{ref}}$
on these same prefixes. It queries $\phi_{\mathrm{AR}}$ with
$M(\mathcal S^\star)$ and $\theta_{\mathrm{ref}}$ under ordinary conditions,
using the support, eligibility, and weighting rules of
Appendix~\ref{app:opsd-planning}.

\textsc{RolloutAndQuery} compiles the student plan, runs the student
renderer, and selects state--time pairs $(Z_j,t_j)$ for teacher queries.
We use $R=3$ independent noise trajectories, each with $K=20$ Euler
steps and guidance 1.0.
Each trajectory starts at $t=1$ and ends at $t=0$; its 20 velocity
evaluations occur at $t=1,0.95,\ldots,0.05$.
We partition these evaluations into four consecutive strata of five
calls and sample one call uniformly from each stratum.
The pooled set $\mathcal Q$ therefore contains $4R=12$ queries, indexed
by $j$, with equal weights $\omega_j=1/(4R)=1/12$.
The terminal state at $t=0$ is excluded.

The three losses follow Equations~\ref{eq:opsd-local-teacher},
\ref{eq:opsd-terminal-flow}, and~\ref{eq:opsd-reference-retention}.
Here $\operatorname{sg}$ denotes stop-gradient: plans, sampled states,
and teacher targets remain detached, while student predictions and
trainable condition encoders are recomputed with gradients.
\textsc{AdamWStep} performs one AdamW update after clipping the
minibatch-averaged gradient at norm 1, using the settings in
Appendix~\ref{app:opsd-optimization}.

\paragraph{Clean prediction matching.}
Let $N=\|m\|_1>0$ count valid latent coordinates under the binary mask.
At a student query $Z_j$, let $v_j^S,v_j^T$ denote the student and
target-conditioned teacher velocities in Algorithm~\ref{alg:joint-opsd},
and $e_j=\|m\odot(v_j^S-v_j^T)\|_2/\sqrt N$.
For $b\in\{S,T\}$, the $t_j^2$ weight in
Equation~\ref{eq:opsd-terminal-flow} follows from
\begin{equation}
 \widehat z_{0,j}^{b}=Z_j-t_jv_j^b,
 \qquad
 \frac{1}{N}\|m\odot(\widehat z_{0,j}^{S}
 -\operatorname{sg}(\widehat z_{0,j}^{T}))\|_2^2
 =t_j^2e_j^2.
 \label{eq:opsd-clean-latent-equivalence}
\end{equation}

\paragraph{Finite-step teacher tracking.}
Fix the student plan, source latents, teacher tokens, and both models'
parameters. For one noise trajectory, index Euler states by
$j=0,\ldots,K$, with $t_j=1-j/K$ and step $h=1/K$.
Compare student and teacher trajectories $Z_j^S,Z_j^T$ from the same
initial noise and with fixed zero padding.
Assume teacher velocities are $L$-Lipschitz on valid coordinates over
a domain containing both trajectories at each grid time $t_j>0$.
Define $\delta_j=\|m\odot(Z_j^S-Z_j^T)\|_2/\sqrt N$ and
$\ell_{\mathrm{full}}=K^{-1}\sum_{j=0}^{K-1}t_j^2e_j^2$,
evaluating both velocities in $e_j$ at $Z_j^S$. Then
\begin{equation}
 \delta_K^2\le C_{K,L}\ell_{\mathrm{full}},
 \qquad
 C_{K,L}=Kh^2\sum_{j=0}^{K-1}
 \frac{(1+hL)^{2(K-1-j)}}{t_j^2}.
 \label{eq:opsd-trajectory-bound}
\end{equation}
Indeed, the backward-time Euler update subtracts $h$ times the velocity.
Adding and subtracting the teacher velocity at $Z_j^S$ gives
$\delta_{j+1}\le(1+hL)\delta_j+he_j$.
With $\delta_0=0$, unrolling yields
$\delta_K\le h\sum_{j=0}^{K-1}(1+hL)^{K-1-j}e_j$;
weighted Cauchy--Schwarz gives Equation~\ref{eq:opsd-trajectory-bound}.
For $R$ collected trajectories, let $\ell_{\mathrm{full}}^{(r)}$
denote the full-grid loss on trajectory $r$.
Uniform querying within the four equally sized strata, with weights
$1/(4R)$, gives
$\mathbb E_{\mathcal Q}[\ell_{\mathrm{DiT}}\mid\{Z_{r,j}^S\}]
=R^{-1}\sum_{r=1}^{R}\ell_{\mathrm{full}}^{(r)}$.
Thus the sampled objective estimates the full-grid mismatch entering
the bound, linking local supervision at student states to finite-step
tracking of the prepared teacher under the executed plan.
The teacher trajectory serves as an analytical comparison; training
queries the teacher at student states, while target-audio accuracy
depends on teacher quality.


\subsection{Shared gradients and supervision variants}
\label{app:opsd-optimization}

\paragraph{Joint OPSD.}
All three losses in Equation~\ref{eq:opsd-objective} update shared
parameters and their active task modules.
Student refinement adds no paired GT token, flow, or structured loss
$\mathcal L_{\mathrm{struct}}$.
GT plans supply teacher memory and eligibility facts; GT audio supplies
teacher reference tokens. Paired teacher preparation is defined in
Appendix~\ref{app:opsd-teachers}.

\paragraph{Branch supervision variants.}
\label{app:opsd-variants}
The two ablations retain the shared student and replace one distillation
 term with its GT counterpart:
\begin{equation}
 \mathcal L_{\mathrm{AR\ OPSD}}
 =\mathbb E_{\mathcal D}\!\left[
 \lambda_{\mathrm{AR}}\ell_{\mathrm{AR}}
 +\lambda_{\mathrm{DiT}}\ell_{\mathrm{RF}}^{\mathrm{GT}}
 +\lambda_{\mathrm{ref}}\ell_{\mathrm{ref}}\right],\quad
 \mathcal L_{\mathrm{DiT\ OPSD}}
 =\mathbb E_{\mathcal D}\!\left[
 \lambda_{\mathrm{AR}}\ell_{\mathrm{CE}}^{\mathrm{GT}}
 +\lambda_{\mathrm{DiT}}\ell_{\mathrm{DiT}}
 +\lambda_{\mathrm{ref}}\ell_{\mathrm{ref}}\right].
 \label{eq:opsd-branch-variants}
\end{equation}
\textbf{AR OPSD} keeps planning distillation on student prefixes.
Its paired flow term $\ell_{\mathrm{RF}}^{\mathrm{GT}}$ uses GT target
plans $\mathcal S^\star$, source latents $z_s$, and interpolated states
$\zeta_t=(1-t)z_y+t\epsilon$ with velocity targets $\epsilon-z_y$
(Equation~\ref{eq:rf}); the audio reference teacher is unused.
\\
\textbf{DiT OPSD} uses next-token cross-entropy
$\ell_{\mathrm{CE}}^{\mathrm{GT}}$, averaged over valid GT tokens with
GT prefixes and ordinary editing conditions; the AR teacher is unused.
Rendering retains joint OPSD's student plans, visited states, and
teacher with GT audio reference tokens. GT prefixes enter only the token loss; DiT rollouts use student plans.

Both variants retain the reference model, field selection, eligibility
rules, and $\lambda_{\mathrm{ref}}$ of joint OPSD; neither adds
$\mathcal L_{\mathrm{struct}}$.
Both modes remain trainable. Shared updates let either loss affect both modes in later rollouts.

\paragraph{Optimization.}
The three OPSD configurations use the same student initialization,
native parameter groups, and global batch of 64 requests, with
$\lambda_{\mathrm{AR}}=0.1$, $\lambda_{\mathrm{DiT}}=1$, and
$\lambda_{\mathrm{ref}}=0.1$.
Learning rates are $5\times10^{-6}$ for AR adapters and structured heads
and approximately $3.75\times10^{-7}$ for shared and rendering parameters.
Shared weights occur once in AdamW. Request losses accumulate before
global averaging, gradient clipping at 1, and one optimizer step.
Unavailable rollout losses contribute zero without changing the batch
denominator; GT terms remain defined from the paired record.
Updates use fresh rollouts, with teachers, the reference model, text
backbones, CLAP, and VAE frozen.

\subsection{Post-training controls and comparison protocol}
\label{app:post-training-controls}

Table~\ref{tab:opsd-editing} compares six post-training configurations:
continued SFT, ReFL, GRPO, AR OPSD, DiT OPSD, and joint OPSD.
All retain the native shared and task specific parameter groups, with
text backbones, \method{} CLAP, and VAE frozen.
The OPSD variants follow Appendix~\ref{app:opsd-optimization};
$\mathcal L_{\mathrm{paired}}$ below denotes Equation~\ref{eq:edit-joint},
including structured supervision.

\paragraph{Comparison protocol.}
All recipes use identical test requests, noise seeds, and sampling
settings for evaluation. Their paired batches, reward rollouts, trajectory
reuse, and teacher preparation follow the settings specified below.

\paragraph{SFT.}
SFT continues $\mathcal L_{\mathrm{paired}}$ with 192 paired examples
per update and no reward objective.

\paragraph{ReFL.}
Our ReFL adaptation rewards frozen AMBIT CLAP semantic similarity between
the generated terminal latent $\hat z_0$ and raw edit instruction $e$:
\begin{equation}
 \mathcal L_{\mathrm{ReFL}}
 =\mathcal L_{\mathrm{paired}}
 -\lambda_{\mathrm{ReFL}}\,
 \mathbb E\!\left[
 \cos\!\left(f_{\mathrm{FOA}}(\widehat z_0),f_{\mathrm{text}}(e)\right)
 \right],
 \label{eq:post-training-controls}
\end{equation}
with $\lambda_{\mathrm{ReFL}}=1$, 144 paired examples, and three reward
requests per update. Plans are sampled without gradients; reward
backpropagation traverses the final four of 20 Euler steps with
guidance 1.0, updating rendering and shared parameters.
Paired supervision also updates AR; detaching the plan does not freeze
the planner's shared representation.

\paragraph{Reward scope.}
ReFL and GRPO use the semantic head, which treats content-preserving
spatial edits as positives (Section~\ref{sec:transfusion}).
Their reward scores instruction compatibility without directly evaluating
direction, timing, or unedited-source preservation; a relative instruction
also need not describe the full target scene.
Both retain paired spatial and preservation supervision, while their
reward lacks the target plan and audio information supplied to OPSD teachers.

\paragraph{GRPO.}
The two-branch control uses categorical policy
optimization~\citep{shao2024deepseekmath} and Flow-GRPO
transitions~\citep{liu2026flow}.
Each request samples $G=6$ plans and trajectories using the ReFL reward;
undecodable plans receive $-1$ and no trajectory.
Rewards $r_j$ give detached group-normalized advantages:
\begin{equation}
 A_j=\frac{r_j-\bar r}
 {\sqrt{G^{-1}\sum_{k=1}^{G}(r_k-\bar r)^2}+\varepsilon},
 \qquad \bar r=\frac{1}{G}\sum_{k=1}^{G}r_k,
 \label{eq:grpo-group-advantage}
\end{equation}
With $\varepsilon=10^{-6}$, AR samples legal tokens at temperature 0.8.
DiT uses 20 Gaussian SDE steps from noise ($t=1$), with exploration scale 0.3.

For branch $b\in\{\mathrm{AR},\mathrm{DiT}\}$, define
$\rho_{ju}^b=\exp(s_{\theta,ju}^b-\operatorname{sg}(s_{\mathrm{old},ju}^b))$.
AR scores are categorical log probabilities; DiT scores average Gaussian
transition log densities over valid latent coordinates, yielding a
dimension-normalized surrogate instead of a full-vector likelihood ratio.
The clipped objective is
\begin{equation}
 \mathcal L_{\mathrm{clip}}^b
 =-\frac{1}{G}\sum_{j=1}^{G}
 \operatorname{avg}_{u\in\mathcal U_j^b}
 \min\!\left(\rho_{ju}^bA_j,
 \operatorname{clip}(\rho_{ju}^b,1-\delta,1+\delta)A_j\right),
 \label{eq:grpo-clipped-objective}
\end{equation}
with $\delta=0.2$.
Each candidate uniformly queries up to eight non-forced AR decisions
and two DiT transitions; an empty DiT set contributes zero.
Each collected behavior group is reused for two successive optimizer updates.

GRPO retains paired supervision and frozen reference penalties:
\begin{equation}
 \mathcal L_{\mathrm{GRPO}}
 =\mathcal L_{\mathrm{paired}}
 +\sum_{b\in\{\mathrm{AR},\mathrm{DiT}\}}
 \left[\lambda_{\mathrm{G},b}\mathcal L_{\mathrm{clip}}^b
 +\lambda_{\mathrm{KL},b}\mathcal K_b\right].
 \label{eq:grpo-total-objective}
\end{equation}
Here $\mathcal K_{\mathrm{AR}}$ averages
$D_{\mathrm{KL}}(p_\theta\|p_{\mathrm{ref}})$ on legal support;
$\mathcal K_{\mathrm{DiT}}$ averages
\mbox{$\|m\odot(\mu_\theta-\mu_{\mathrm{ref}})\|_2^2/(2\sigma^2\|m\|_1)$},
the equal-variance Gaussian KL averaged over valid coordinates.
Both branches use $\lambda_{\mathrm{G},b}=1$ and
$\lambda_{\mathrm{KL},b}=0.02$, alongside 144 paired examples per update.
Action scores are recomputed with gradients; sampled actions, states,
rewards, advantages, and reference predictions are detached.
\endgroup

\end{document}